\documentclass{aa} % for a paper on 1 column  

\pdfmapfile{+txfonts.map}

\usepackage{txfonts}
\usepackage{graphicx}
\usepackage{color}
\usepackage{natbib}
\usepackage{array}
\usepackage{subfigure}
\usepackage{enumitem}
\usepackage{float}

\begin{document}

\title{Turbulent gas accretion between supermassive black holes and star-forming rings in the circumnuclear disk }
\titlerunning{Turbulent gas accretion between supermassive black holes and star-forming rings in the circumnuclear disk} 

\author
  {Wara Chamani
  \inst{1,2}
  \and
 Stephanie D\"{o}rschner
  \inst{3}
  \and 
Dominik R.\,G. Schleicher
  \inst{4}
  }

\institute{Aalto University Metsähovi Radio Observatory, Metsähovintie 114, 02540 Kylmälä, Finland \and
 Aalto University Department of Electronics and Nanoengineering, P.O. Box 15500, 00076 Aalto, Finland \\
\email{wara.chamani@aalto.fi}
\and
Institut f\"ur Astrophysik, Georg-August-Universit\"at G\"ottingen, Friedrich-Hund-Platz 1, D-37077 G\"ottingen, Germany 
%\email{doerschner@astro.physik.uni-goettingen.de}
\and
Departamento de Astronom\'{i}a, Facultad Ciencias F\'{i}sicas y Matem\'{a}ticas, Universidad de Concepci\'{o}n, 
Av. Esteban Iturra s/n Barrio Universitario, Casilla 160-C Concepci\'on, Chile\\
%\email{dschleicher@astro-udec.cl}
}

\date{Received 16.01.2017/ Accepted 14.03.2017}

%\maketitle

\abstract
% {context.}
% {Aims.}
% {methods.}
% {REsults}
% {Conclusions.}
{While supermassive black holes are known to co-evolve with their host galaxy, the precise nature and origin of this co-evolution is not clear. 
We here explore the possible connection between star formation and black hole growth in the circumnuclear disk (CND) to probe this connection in the vicinity close 
to the black hole. We adopt here the circumnuclear disk model developed by  \citet{Kawakatu:2008rg} and \citet{Wutschik:2013caa}, and explore both the dependence on the star formation recipe 
as well as the role of the gravitational field, which can be dominated by the central black hole, the CND itself or the host galaxy. A specific emphasis is put on the 
turbulence regulated star formation model by  \citet{Krumholz:2005mt} to explore the impact of a realistic star formation recipe. It is shown that this model helps to introduce
realistic fluctuations in the black hole and star formation rate, without overestimating them. Consistent with previous works, we show that the final black hole masses are 
rather insensitive to the masses of the initial seeds, even for seed masses of up to $10^6$~M$_\odot$. In addition, we apply our model to the formation of high-redshift 
quasars, as well as to the nearby system NGC~6951, where a tentative comparison is made in spite of the presence of a bar in the galaxy. We show that our model 
can reproduce the high black hole masses of the high-redshift quasars within a sufficiently short time, provided a high mass supply rate from the host galaxy. In addition, it 
reproduces several of the properties observed in NGC~6951. With respect to the latter system, our analysis suggests that supernova feedback may be important to create 
the observed fluctuations in the star formation history as a result of negative feedback effects. }


\maketitle

% \end{abstract}

\section{Introduction}
The processes which regulate the co-evolution of supermassive black holes (SMBH) with their host galaxies are still not  well understood, as well as the activity 
of SMBH over cosmological time scales. However, SMBH  have been successfully  detected both directly and indirectly at the centers of the 
galaxies, and their masses span from millions to billions of solar masses \citep{Fan2004, Fan2006, Mortlock:2011lo, Venemans2012ApJ, Willot2003ApJ}. A widely accepted 
view suggests that the growth of SMBH occurs due to rapid accretion and merging. However,  it is still unclear what physical  mechanisms  drive SMBH to  reach such
high masses  and what gives rise to the correlation with  the  physical properties of their host galaxies.   
 
Substantial efforts were undertaken and numerous models have been proposed to explain the origin and formation of SMBH through different mechanisms \citep{Rees:1984tin}.
However, to explain the formation of SMBH at higher redshift requires to study the first  black hole (BH) seeds. So far,  there are three main scenarios which  include the 
formation of BH seeds through the core collapse of Population III stars \citep{Trenti07}, the collapse of  dense stellar clusters \citep{Devecchi10, Devecchi12} and the
collapse of protogalactic metal free gas clouds in primodial halos \citep[see e.g.][]{Latif2016}. 
The latter mechanism provides the potentially largest seed masses, and may thus favor the formation of supermassive black holes 
at $z>6$ \citep[e.g.][]{Lodato07, Wise08, Begelman09, Schleicher:2010apj, Latif:2013dua, Latif:2013pyq, Schleicher13, Ferrara14}.
These high-redshift black holes can potentially be detected with ALMA using CO and H2 line emission (\cite{Spaans2008}; \cite{Schleicher:2010AA}).

Several studies have found tight correlations between black hole mass and galaxy bulge properties in the nearby universe. Such
correlations have indicated scaling relations of black hole mass with bulge mass as well as black hole mass with the velocity dispersions of their host 
(\cite{Magorrian:1997hw}; \cite{Haring:2004hr} \cite{Marconi:2003hj}; \cite{Merritt:1999ry}; \cite{Gebhardt:2000fk};\cite{Ferrarese:2000se}).
Correlations between star formation rate and SMBH growth have also been found by \citet{DiamondStanic:2011md}. Their work includes a complete 
sample of Seyfert galaxies and they found a strong correlation between nuclear ($r<1$ kpc)  star formation rates  and the accretion rates of the SMBH 
with a scaling relation $\dot{M}_{*}\sim \dot{M}^{0.8}_{\rm{BH}}$. Measurements on scales $r>1$ kpc with extended star formation in the host galaxy showed only a 
weak correlation and  scaling as $M_{*}\sim \dot{M}^{0.6}_{\rm{BH}}$. Their results suggest a connection between gas on sub-kiloparsec scales 
that is forming stars and the gas on sub-parsec scales that goes into black hole accretion. In this case, the transport of angular momentum in the galactic 
disk would be a natural mechanism which plays a significant role in feeding the central engine and triggering star  formation. Additionally, prominent circumnuclear  
disks with star formation rings have been observed in the inner regions of some galaxies from parsec scale up to a few kilo parsecs from the galactic 
center \citep{Lenc:2009le,Hsieh:2011ki,vanderLaan:2013qy,falcon:2013}.
 
 In the last years for several active galaxies gas mass inflows to the nucleus have  been detected through optical and near-infrared spectroscopy with estimated
 inflow rates of the order of $10^{-2} M_{\odot} yr^{-1}$ \citep{Schnorr2016MNRAS, Lena2015ApJ, Schnorr2014MNRASa, Schnorr2014MNRASb, Storchi2009MNRAS, Storchi2010MNRAS}. 
 However, from the current observations it is difficult to determine the actual kinematics of the inflowing gas to the nucleus \citep{Storchi2010MNRAS}.

 Co-evolution models and numerical simulations are necessary for explaining and describing the processes that regulate and control the accretion onto SMBH 
 as well as the gas inflow  from the host galaxy which triggers nuclear starbursts from kilo-parsec scale down to parsec scale in active galaxies.
 \citet{Kawakatu:2008rg} and \citet{Wutschik:2013caa} have proposed models of co-evolution,  where  the transport of angular momentum from the outer border of the CND to 
 the SMBH is driven by turbulence injected via supernova explosions that lead to different accretion phases in the disk around a SMBH. An important aspect 
 studied by \citet{Wutschik:2013caa} is the functional form of the star formation rate, as different dependencies have been suggested in the literature with respect 
 to the surface density of the gas as well as the turbulent velocity. The models explored included both linear and nonlinear dependencies on the gas surface density, as well as
 models independent of the turbulent velocity $\rm{v}_{turb}$ and models where the star formation rate scales as $\rm{v}_{turb}^{-1}$. Specific models adopted from the literature
 included in particular the model employed by \citet{Kawakatu:2008rg} and by \citet{Elmegreen:2009mv} (hereafter model EB10) .
 
In this work, we further extend the framework of \citet{Wutschik:2013caa} and apply it to the formation of both nearby AGN and high-redshift quasars. For this purpose, 
we include the star formation recipe by \citet{Krumholz:2005mt} (hereafter model KM05), including both non-linear dependencies on gas surface density and turbulent
velocity, from which the Kennicutt-Schmidt relation can be derived in a systematic manner. We also explore whether our assumptions regarding the gravitational dynamics 
have a strong impact on the results. With this new model, we will show that the non-linear dependence in the star formation model, along with feedback from supernova 
explosions, can give rise to star formation histories that are similar to the observed ones. 

In section 2, we describe the general implementation along with the new star formation model considered here. Section 3 shows the results of semi-analytical calculations
and simulations which describe the time evolution of the  black hole mass, stellar mass and disk radius, exploring different assumptions on the dominant component in 
the gravitational field. We also show  the time evolution of the Mach numbers and the expected evolution of observable quantities as the luminosities. In section 4, the 
model is then applied to specific objects including local AGN and high-redshift quasars. Finally section 5 presents our discussion and conclusions.

\section{Evolution of the SMBH and star formation in a circumnuclear disk}
%2.2 Description of the star formation model	
\subsection{Disk Model}
 
The co-evolution model of \citet{Kawakatu:2008rg} considers the following system:  an accreting SMBH surrounded by a CND, all embedded in the galaxy. 
The host galaxy supplies matter (dusty gas) with the rate $\dot{M}_{sup}$ to the CND. The disk becomes gravitationally unstable at a critical radius $r_{c}$, 
 where the Toomre Q parameter is equal to 1. This separates the disk into an outer disk  (self-gravitating region) and into an  inner disk (stable region). 
 In the gravitationally unstable outer disk stars are forming and  the turbulent pressure produced by supernova explosions drives gas to the inner disk to feed the SMBH. 

In the extension of the model by \citet{Wutschik:2013caa} (hereafter model WS13), the stable and the unstable part of the CND are treated separately, 
assuming a power-law surface density distribution where the power-law index is obtained from normalization constraints. The mass exchange between both 
parts of the disk is driven by direct mass exchange due to accretion from the unstable outer disk to the inner disk, as well as through geometric effects, i.e. if the 
critical radius moves towards the inner or outer part of the disk. The model also includes the cases where the entire CND is stable or unstable, i.e. where the critical radius
is equal to the inner or outer radius of the CND (see also  \citet{Kawakatu:2008rg}). \citet{Wutschik:2013caa} and  \citet{Kawakatu:2008rg} show that 
depending on the stability of the CND, one can distinguish between the following phases:\\

a) High accretion phase (constant supply from the host galaxy):
\begin{itemize}
 \item  Due to higher matter density or lower turbulent velocity, stable matter  in the disk becomes unstable , which results into the critical radius moving inward,
  i.e. $r_{c}(t_{o})> r_{c}(t)$ or $r_{c}\leq r_{in}$.
   \item In consequence, gas $\Delta M_{g}$ from the inner gas reservoir flows to the outer reservoir. Hence the gas mass in the outer region $M_{g,o}$ is defined as
  $M_{g,o}(t)=M_{g,o}(t_{o})-\Delta M_{g}$.
 \item The disk becomes geometrically thick, stars form and their feedback through supernova explosions leads to the turbulent supersonic phase. In this phase, the disk thickness is regulated by the turbulent pressure  of the gas $P_{g}$ which is defined as $P_{g}=\rho_{g}v^{2}_{turb}$, where $\rho_{g}$ is the gas density and the turbulent velocity $v_{turb}$ is generally larger  than the sound speed $c_{s}$, i.e. $v_{turb}> c_{s}$. 
 \item The turbulent energy in the disk is $E_{turb}(r)=\Sigma_{g}v^{2}_{turb}(r)$, where $\Sigma_{g}$ is the surface density of the gas. 
   \item  The total energy injected by supernovae is given as $\dot{E}_{inj}(r,t)= f_{SN} \eta_{SN} E_{SN} \hat{{{\dot{\Sigma}}}}_{*}(r,t-t_{SN})$, with $t_{SN}$ is the average  
   life-time of massive stars which then explode into supernovae, $E_{SN}$ the energy injected into the ISM via  supernova, $\eta_{SN}$ the heating efficiency. and
   $f_{SN}$ is a constant parameter.  $\hat{{{\dot{\Sigma}}}}_{*}(r,t-t_{SN})$ is the local star formation rate per unit area.  We
   assume here that the energy not going into turbulence is dissipated ($E_{dis}$) through other processes, and due to the energy balance, we then 
   have $\dot{E}_{turb}(r)+\dot{E}_{dis}(r)=\dot{E}_{inj}$. As a typical lifetime for massive stars, we adopt $10^6$~years, but we will also explore variations of this parameter in section 4.
  \item The feedback from supernova explosions is the primary mechanism which drives the gas to the inner disk to feed the black hole. During this phase,  accretion becomes highly efficient.
 \end{itemize}

 b) Low accretion phase (after termination of supply from host galaxy):
 \begin{itemize}
  %\item The circumnuclear disk is fully gravitationally stable.
  \item Gravitationally unstable matter in the disk (in a state of active star formation) becomes stable which results into the critical
  radius moving outward, i.e. $r_{c}(t_{o})< r_{c}(t)$ or  $r_{c}>r_{in}$.
  \item The gas reservoir of the inner stable disk therefore increases, as a larger part of the CND is becoming stable over time. 
  \item Once stable, the disk is vertically supported only by thermal gas pressure, $P_{g}=\rho_{g}c^{2}_{s}$, where $v_{turb}<c_{s}$.
  \item The scale height of the disk is determined by the balance between gravity and thermal gas pressure. Accretion is inefficient in this phase.
 \end{itemize}
 
 In our model we employ  the input parameters for the outer and inner radius of the disk as  $r_{out}=\sqrt{M_{BH}/\pi \Sigma_{host}}$
and $r_{in}=3 pc \sqrt{M_{BH}/10^{8}M_{\odot}}$ respectively \citep{Wutschik:2013caa, Kawakatu:2008rg}. Table ~\ref{tab:1} summarizes the fixed and free parameters 
of our model. The values for the time of supply $t_{sup}$,  the surface density of the host galaxy $\Sigma_{host}$,  the power law exponent for the surface gas
density $\gamma$, the parameters related to the energy injected by supernova explosions  ($t_{SN}$, $f_{SN}$, $\eta_{SN}$ and $E_{SN}$), and the thermal sound 
speed $c_{s}$ are taken from previous works  \citep{Wutschik:2013caa, Kawakatu:2008rg}. Considering  mass conservation, the mass balance in the system
is   $M_{sup}=M_{BH}+M_{g}+M_{\ast}$, where  $M_{sup}$ is the mass supplied to the CND, $M_{BH}$ is the black hole mass, $M_{g}$ is the gas mass and  $M_{\ast}$  is  
the stellar mass. The time evolution of the gas mass in the disk is given as  
 \begin{equation}
 \label{eq:1}
 M_{g}(t)=\int_{0}^{t} [\dot{M}_{sup}(t')-\dot{M}_{*}(t')-\dot{M}_{BH}(t')]\,dt' ,
\end{equation}
considering the difference between the mass that was supplied and the mass going into stars and the central black hole. $\dot{M}_{sup}$ is the mass supply rate, 
$\dot{M}_{BH}$ is the black hole accretion rate and  $\dot{M}_{*}$ is the star formation rate in the disk. The latter is  given by the following expression:
\begin{equation}
\label{eq:2}
 \dot{M}_{*}(t)=\int_{r_{min}}^{r_{out}} 2 \pi r \dot{\Sigma}_{*}(r)\,dr ,
\end{equation}
where the value of $r_{min}$ depends also on the disk stability and is given by $r_{min}={\rm max}(r_{in}, r_c)$. To estimate the black hole mass accretion rate in a viscous accretion disk, we refer to the expression  given by  \citet{Pringle:1981ds} as

\begin{equation}
\label{fig:3}
\dot{M}_{BH}(r,t)=2\pi\nu\Sigma_{g}(r) \frac{d\ln\Omega(r)}{d\ln r},
\end{equation}

where $\nu = \alpha {\rm{v}(r)} h(r)$ denotes the viscosity, with $\alpha$ the viscous coefficient, $\rm{v}(r)$ the gas velocity, and $h(r)$ 
the scale height of the disk.  Here $\Omega$ represents the angular velocity of the gas, which is obtained from the balance between the centrifugal force and the gravitational force. The latter depends on the dominant gravitational component, which can be the black hole, the CND itself and the host galaxy. In the general case, it is thus given as
\begin{equation}
\label{eq:4}
\Omega^{2}(r)=\frac{GM_{BH}}{r^{3}}+\frac{\pi G} {r}(\Sigma_{disk}(r)+\Sigma_{host}),
\end{equation}
where $\Sigma_{host}$ is the host galaxy surface density, $\Sigma_{disk}=\Sigma_{g}+ \Sigma_{stars}$ is 
the disk surface density. It follows a power law function of the radius, represented as $\Sigma_{disk}\approx \Sigma_{g}= \Sigma_{0}(r/r_{c})^{-\gamma}$, where $\gamma$ is a free parameter \citep{Wutschik:2013caa}. 
The evolution of the black hole mass $M_{BH}$ from an initial seed mass $M_{BH,seed}$ is  given as
\begin{equation}
\label{eq:5}
 M_{BH}(t)=M_{BH,seed}+\int_{0}^{t} \dot{M}_{BH}(t')\,dt'.
\end{equation}

 \begin{table}[t!]
  \centering
  \caption{Model parameters.}
 % \vspace{0.3cm}
  %\begin{tabular}{lll}
  \begin{tabular}{  m{1cm}  m{4.1 cm} m{1.9cm} m{ 1cm}  } 
    \hline
  	 Type  & Name & Values & fix/free\\
	\hline\noalign{\smallskip}
	%\multirow{2}{*}{Defenders} 
	$\dot{M}_{sup}$  &	Mass supply rate &  1 $M_{\odot}$ $\rm{yr}^{-1}$ &free \\ 
	$t_{sup}$	 &	Time supply in years & $10^{8}$ yr & free\\
	$t_{max}$ & Maximum time for iteration &  $10^{9}$ yr & free\\
        $\Sigma_{host}$	 &	Host surface density of gas  & $10^{4}$ $M_{\odot}$ $\rm{pc}^{-2}$ &free \\
        $\alpha$&Viscosity coefficient& 1 & free\\
	 $\gamma$&Gas surface density power law exponent&1 &fix\\
        $M_{BH,seed}$	 &	Black hole seed & $10^{3}$ $M_{\odot}$& free \\ 
        $t_{SN}$ &Average life time of massive stars&$\approx10^{6}$ yr& free\\
        $f_{SN}$&Supernovae rate per solar mass of formed stars&$7.9\times 10^{-3}$& fix\\
        $\eta_{SN}$&supernova heating efficiency&0.1&fix\\
        $E_{SN}$&Thermal energy by core-collapse supernovae&$10^{51}$ erg/s&fix\\
        $c_{s}$		 &	Sound speed& 1 km/s&fix\\
        $f_{\rm{GMC}}$ 	 &	Fraction of molecular clouds & 0.5& fix \\
         $Q$		 &	Toomre parameter& 1 &fix\\
        $\Psi$		 &	Dependent on $f_{\rm{GMC}}$ and $Q$& 0.1224 &fix\\
        
	\hline
  \end{tabular}
    \label{tab:1}
\end{table}

\subsection{Star formation rate model}

As a new ingredient in our model, we consider here the star formation model of KM05 due to its effective 
applicability  not only to the Milky Way's star formation rate but also to a vast variety of galaxies and  its consistency with observations and simulations. 
The model particularly allows to derive the two forms of the Kennicutt-Schmidt relation and successfully predicts star formation rates based on observable 
quantities over a large range of conditions, spanning from normal galactic disks, circumnuclear starburts and ULIRGs to nearby star-forming galaxies such as 
Markarian~273, Arp~193 and Arp~220.

In the  analytical description of KM05, the main premise is  that the gas collapses in  sub-regions of molecular
clouds in a supersonically turbulent environment to form stars. The total  star formation rate $\dot{\Sigma}_{*}$  is formulated in terms of the observable properties 
of the gas in the galactic disk as follows

\begin{equation}
\label{eq:6}
\dot{\Sigma}_{*} \approx 0.073 \textit{M}^{-0.32} \phi^{1/2}_{\rho} Q^{-1}_{1.5} f_{\rm{GMC}} \Omega \Sigma_{g}\left[ M_{\odot} \rm{pc^{-2}yr^{-1}}\right],
 \end{equation}
 
where $\textit{M}$ is the mach number which is the ratio of the turbulent velocity $\rm{v}_{turb}$ to the local  sound speed $c_{s}$.
 The observable parameter $\phi_{\rho}$ is given by the ratio of the mean molecular density to the 
mean midplane density. This  ratio  has been approximated as $5(\phi_{\bar{P}}/6)^{3/4}$, where $\phi_{\bar{P}}\approx 10-8f_{\rm{GMC}}$ \citep{Krumholz:2005mt}.
The latter can be expressed as  $\phi_{\bar{P}}\approx 5[\frac{10-8 f_{GMC}}{6}]^{3/4}$, where $f_{\rm{GMC}}$ is the molecular gas fraction.
$f_{\rm{GMC}}=0.5$ (it implies $\phi_{\rho}=0.6$) is considered as an universal parameter which applies from starbursts to normal galaxies, as stated in the appendix  B of \citet{Krumholz:2005mt}. The parameter  $Q$ represents the Toomre parameter \citep{Toomre:1964zx, Rice:2016ri} in equation (\ref{eq:6}),   $Q_{1.5}=Q/1.5$. 

To evaluate the effect of the star formation model on the BH accretion rate  and on the star formation rate evolution in the CND,  
we re-write equation (\ref{eq:6}) by emphasizing the dependence of the star formation rate on the turbulent velocity and angular velocity:

\begin{equation}
\label{eq:7}
\dot{\Sigma}_{*}(r)= \Psi \Omega(r) \left(\frac{\rm{v}_{turb}(r)}{c_{s}}\right) ^{-0.32} \Sigma_{g}(r). 
%\dot{\Sigma}_{*}(r)=0.073\phi^{1/2}_{\rho} Q^{-1}_{1.5} f_{GMC} \Omega(r) \left(\frac{v_{turb}(r)}{c_{s}}\right) ^{-0.32} \Sigma_{g}(r) 
 \end{equation}

We write the fix parameter $\Psi$ in terms of $Q$ and $f_{\rm{GMC}}$ as  $\Psi= 0.073\sqrt{5}\left(\frac{10-8f_{\rm{GMC}}}{6} \right)^{3/8} \dfrac{1.5}{Q}f_{\rm{GMC}}$. 
Now that we have condensed equation (\ref{eq:6}), we  employ in the next section the formulation given in  equation (\ref{eq:7}) for studying  the 
evolution of the  circumnuclear disk. If (part of) the CND is gravitationally unstable, the turbulent velocity in this region will be calculated based on the 
energy input from supernovae, leading to supersonic turbulence. As in the WS13 model, we approximate the turbulent velocity here as a power law via

\begin{equation}
\label{eq:8}
\rm{v}_{turb}={\rm{v}}_{o}\left(\frac{r}{r_{out}}\right)^{\lambda} ,
  \end{equation}

where $\rm{v}_{o}=\sqrt{E_{turb}(r_{c})/\Sigma_{g}(r_{c})}$ and  $\lambda$ is the power law index  expressed as $\lambda=\dfrac{1.5-\gamma(\theta-1)}{(2+\epsilon)}$, 
which is  0 in the sub-sonic turbulence phase.  The parameters  $\theta$ and $\epsilon$ come from the  parametrization of the total star formation rate
as~$\dot{\Sigma}_{*}(r)=\Psi\cdot \left(\Sigma_{g}(r)\right)^{\theta}\cdot(\rm{v}_{turb})^{-\epsilon}$ \citep{Wutschik:2013caa} . 
In the KM05 model the latter two parameters have the values  of 1 and 0.32 respectively (see equation (\ref{eq:7})), hence we use them for calculating the 
value of $\lambda$, which  is approximately equal to 0.65.

  \subsection{Dependence of accretion on dominant gravitational source} 

As described in section 2.2, the total star formation rate depends non-linearly on the turbulent velocity  and linearly on the surface density of gas. 
However, since in equation (\ref{eq:4}) the angular velocity depends on different gravitational source terms, we  insert this relation 
into equation (\ref{eq:7}) to write the total star formation rate  as

\begin{align}
\footnotesize
\label{eq:9}
\dot{\Sigma}_{*}(r) &= \Psi \sqrt{{\frac{GM_{BH}}{r^{3}}+\frac{\pi G} {r} \left( \Sigma_{disk}(r)+\Sigma_{host} \right) }} \cdot  \notag\\
&\phantom{{}=1} \left(\frac{{\rm{v}_{turb}}(r)}{c_{s}}\right) ^{-0.32} \Sigma_{g}(r).
\end{align}

To evaluate the evolution of the star formation rate   $\dot{M}_{*}(t)$  in the CND, one has to insert equation (\ref{eq:9}) into (\ref{eq:2}). To calculate
equation (\ref{eq:2}) analytically, we first consider three different  limiting cases of $\dot{\Sigma}_{\ast}(r)$, depending on the dominant source of gravity in the system: the 
central black hole, the CND itself or the gravitational field from the host galaxy. As a result, we will obtain expressions for the star formation rate $\dot{M}_{*}(t)$ and 
the energy injection rate $\dot{E}_{inj}$ in different regimes. 

If gravity is dominated by the central black hole, we obtain  
$\dot{\Sigma}_{*}(r) = \Psi \sqrt{\left(\frac{GM_{BH}}{r^{3}} \right) } \left(\frac{\rm{v}_{turb}(r)}{c_{s}}\right) ^{-0.32} \Sigma_{g}(r)$, and
the analytic solution for $\dot{M}_{*}(t)$ and  $\dot{E}_{inj}$  is given  as

\begin{align}
\label{eq:10}
 \dot{M}_{*}(t) &= \frac{2\pi \Psi\sqrt{GM_{BH}}}{\frac{1}{2}-\theta \gamma -\epsilon \lambda} \Sigma^{\theta}_{o}\left(\frac{\rm{v}_{o}}{c_{s}}\right)^{-\epsilon}\cdot \notag \\
 &\phantom {{}=1} r^{\theta \gamma +\epsilon \lambda}_{c}  \left[r^{\frac{1}{2}-\theta \gamma -\epsilon \lambda}_{out}-r^{\frac{1}{2}-\theta \gamma -\epsilon \lambda}_{c}\right], \rm{and}
\end{align}

\begin{align}
\label{eq:11}
 \dot{E}_{inj}& =7.9\cdot 10^{-3}\eta_{SN}E_{SN}\frac{ \dot{M}_{*}}{2\pi} \left(\frac{1}{2}-\theta \gamma -\epsilon \lambda \right) \cdot\notag \\
 &\phantom{{}=1} r^{-\theta\gamma-\epsilon\lambda-\frac{3}{2}} \left[r^{\frac{1}{2}-\theta \gamma -\epsilon \lambda}_{out}-r^{\frac{1}{2}-\theta \gamma -\epsilon \lambda}_{in}\right]^{-1}.
\end{align}

If the gravitational force is dominated by the CND itself, then  $\dot{\Sigma}_{*}(r) = \Psi \sqrt{\left(\frac{\pi G \Sigma_{disk}}{r} \right) } \left(\frac{\rm{v}_{turb}(r)}{c_{s}}\right) ^{-0.32} \Sigma_{g}(r)$,
and the analytic solution for $\dot{M}_{*}(t)$ and  $\dot{E}_{inj}$  follows as

\begin{align}
\label{eq:12}
 \dot{M}_{*}(t) &= \frac{2\pi \Psi\sqrt{G\pi\Sigma_{o}}}{\frac{3}{2}-\gamma(\frac{1}{2}+\theta) -\epsilon \lambda} \Sigma^{\theta}_{o}\left(\frac{\rm{v}_{o}}{c_{s}}\right)^{-\epsilon}
 r^{\theta \gamma +\epsilon \lambda+\frac{\gamma}{2}}_{c} \cdot \notag \\
 &\phantom{{}=1} \left[r^{\frac{3}{2}-\gamma(\frac{1}{2}+\theta) -\epsilon \lambda}_{out}-r^{\frac{3}{2}-\gamma(\frac{3}{2}+\theta ) -\epsilon \lambda}_{c}\right], \rm{and}
\end{align}

\begin{align}
\label{eq:13}
 \dot{E}_{inj}& =7.9\cdot 10^{-3}\eta_{SN}E_{SN}\frac{ \dot{M}_{*}}{2\pi} \left(\frac{3}{2}-\gamma(\frac{1}{2}+\theta) -\epsilon \lambda \right) \cdot \notag \\
 &\phantom{{}=1} r^{-\gamma(\frac{1}{2}+\theta)-\epsilon\lambda-\frac{1}{2}} \left[r^{\frac{3}{2}- \gamma(\frac{3}{2}+\theta) -\epsilon \lambda}_{out}-r^{\frac{3}{2}- \gamma(\frac{1}{2}+\theta) -\epsilon \lambda}_{in}\right]^{-1}.
\end{align}

Finally if gravity is dominated by the host galaxy, the star formation rate is expressed as 
$\dot{\Sigma}_{*}(r) = \Psi \sqrt{\left(\frac{\pi G \Sigma_{host}}{r} \right) } \left(\frac{\rm{v}_{turb}(r)}{c_{s}}\right) ^{-0.32} \Sigma_{g}(r)$,
and the analytical solution for $\dot{M}_{*}(t)$ and  $\dot{E}_{inj}$  is given as

 \begin{align}
 \label{eq:14}
 \dot{M}_{*}(t) &= \frac{2\pi \Psi\sqrt{G\pi\Sigma_{host}}}{\frac{3}{2}-\theta \gamma -\epsilon \lambda} \Sigma^{\theta}_{o}\left(\frac{\rm{v}_{o}}{c_{s}}\right)^{-\epsilon}
r^{\theta \gamma +\epsilon \lambda}_{c} \cdot  \notag \\
 &\phantom{{}=1}  \left(r^{\frac{3}{2}-\theta \gamma -\epsilon \lambda}_{out}-r^{\frac{3}{2}-\theta \gamma -\epsilon \lambda}_{c}\right), \rm{and}
\end{align}

\begin{align}
\label{eq:15}
 \dot{E}_{inj}& =7.9\cdot 10^{-3}\eta_{SN}E_{SN}\frac{ \dot{M}_{*}}{2\pi} \left(\frac{3}{2}-\theta \gamma -\epsilon \lambda \right) \cdot \notag \\
 &\phantom{{}=1} r^{-\theta\gamma-\epsilon\lambda-\frac{1}{2}} \left[r^{\frac{3}{2}-\theta \gamma -\epsilon \lambda}_{out}-r^{\frac{3}{2}-\theta \gamma -\epsilon \lambda}_{in}\right]^{-1}.
\end{align}

In the most general case, different regimes (where different terms in the gravitational field may become dominant in different parts of the CND), 
and their evolution may further be time-dependent. As we aim here to work with a still simplified model, we will thus show in the following that the results are 
not highly sensitive to the choice of the gravitational force term. 

 \begin{figure}[t!]
 \centering
 \includegraphics[width=0.5\textwidth]{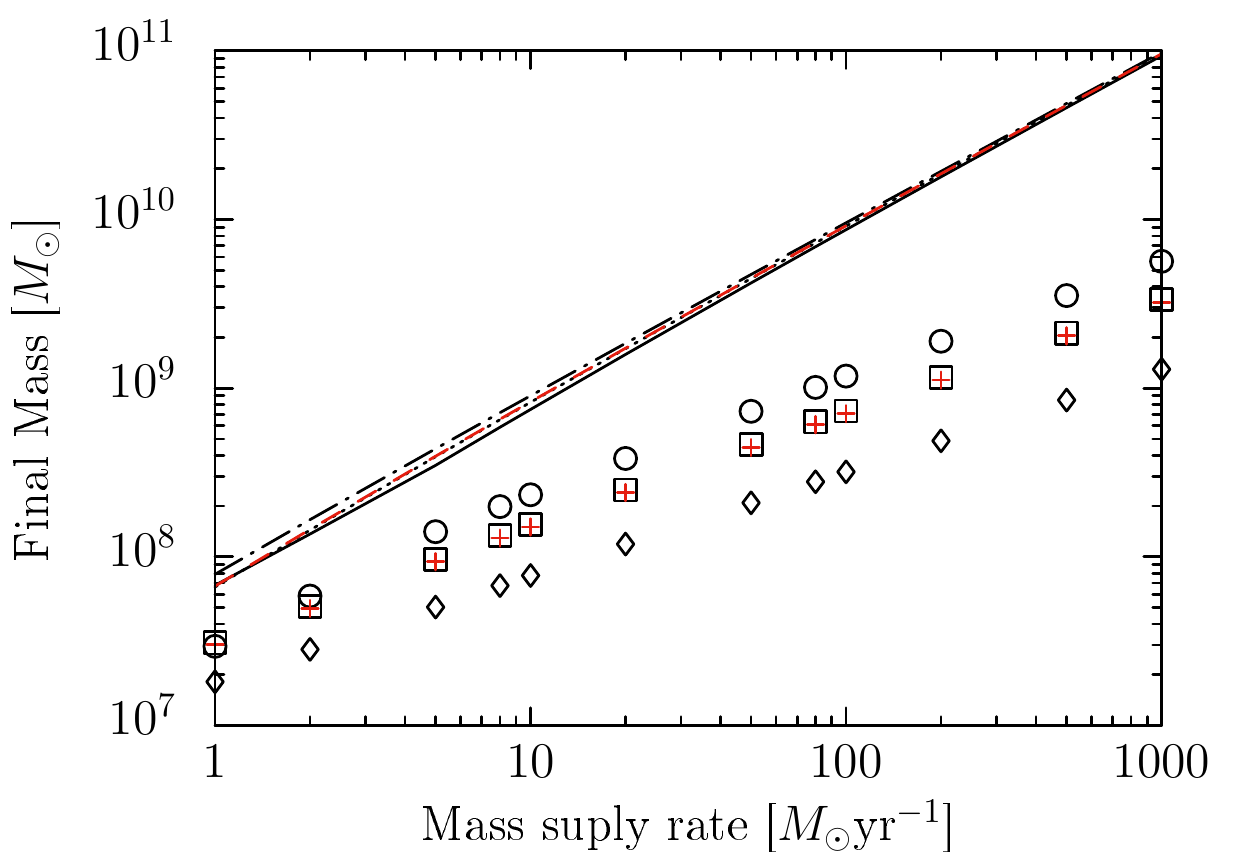}
  \caption{Final black hole (markers) and stellar (lines) masses under different assumptions for the gravitational field, and as a function of 
 the mass supply rate. Results from the KM05 model with the gravitational field dominated by the central black hole (circle and  line), by the CND (square and dotted line),
by the host galaxy (diamond and dotted-dashed line). In addition results for the EB10  model, assuming the CND as the gravitational source (red crosses and red dashed line).}
\label{fig:1}
\end{figure}

\section{Results}

\subsection{Evolution of the SMBH accretion rate, the star formation rate and masses}
%comparisons wiht Stephi's results.
The model described in the previous sections allows us to  explore  the evolution of different physical quantities of the system that include the evolution
of the  black hole growth from an initial seed,  the evolution of  the star formation rate taking place  on the outer disk, 
the evolution of the gas velocity as well as the AGN luminosity. In general, in this model we adopt  a marginally stable self-gravitating 
disk $Q\approx 1$  \citep[consistent with][]{Wutschik:2013caa} and the universal fraction of molecular clouds \cite[$f_{\rm{{GMC}}}=0.5$ as shown by][]{Krumholz:2005mt}.

As mentioned in the previous section, we examine the impact of the star formation model KM05 on the evolution of the system. Figure \ref{fig:1} compares the 
final  black hole mass  and final stellar mass  as a function of the mass supply rate $M_{\odot}$ under different assumptions for the dominant gravitational source, using 
simulations with the  input parameters displayed in table \ref{tab:1}. It can be seen that the final stellar masses are
independent on the assumptions regarding the gravitational field. The black hole mass depends somewhat on the underlying gravitational field, but the difference
amounts to only a factor of a few.  Additionally, in figure \ref{fig:1} we compare our results produced with the KM05 model  with the results produced with the EB10 model.

We note here that by setting  an initial black hole seed of $10^{3}$ $M_{\odot}$ and having a  $\dot{M}_{sup}\geq100$~$M_{\odot}$ $\rm{yr}^{-1}$ from the galaxy to the
circumnuclear disk, both  models KM05 and EB10 enable the growth of supermassive black hole mass to the order of  $10^{9}$~$M_{\odot}$. It is interesting to note that 
the final black hole masses converge to similar values  both in the KM05 model and the EB10 model, here explored in the limit where the CND dominates the gravitational field. 

\begin{figure*}
\begin{minipage}{.5\linewidth}
  \includegraphics[width=\columnwidth]{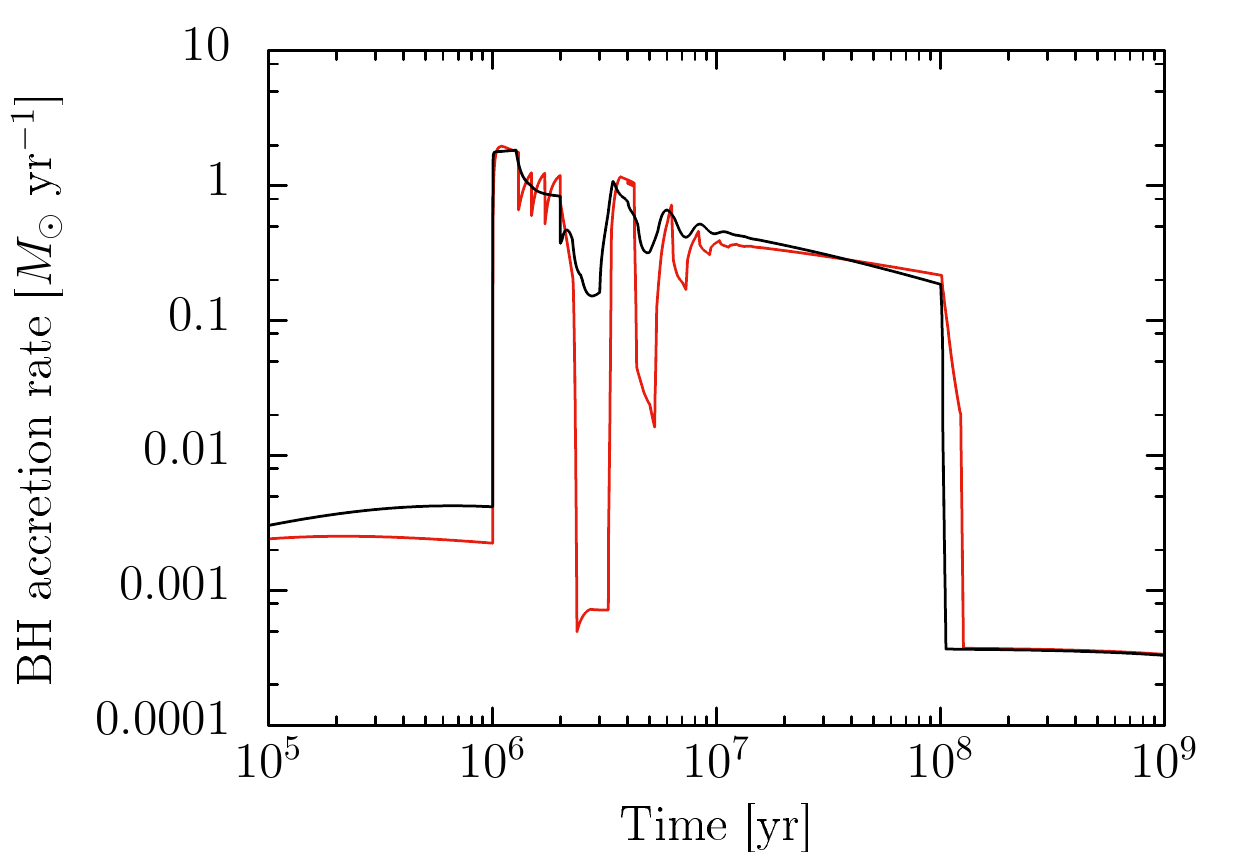}
      %\caption{a}
\end{minipage}
\begin{minipage}{.5\linewidth}
 \includegraphics[width=\columnwidth]{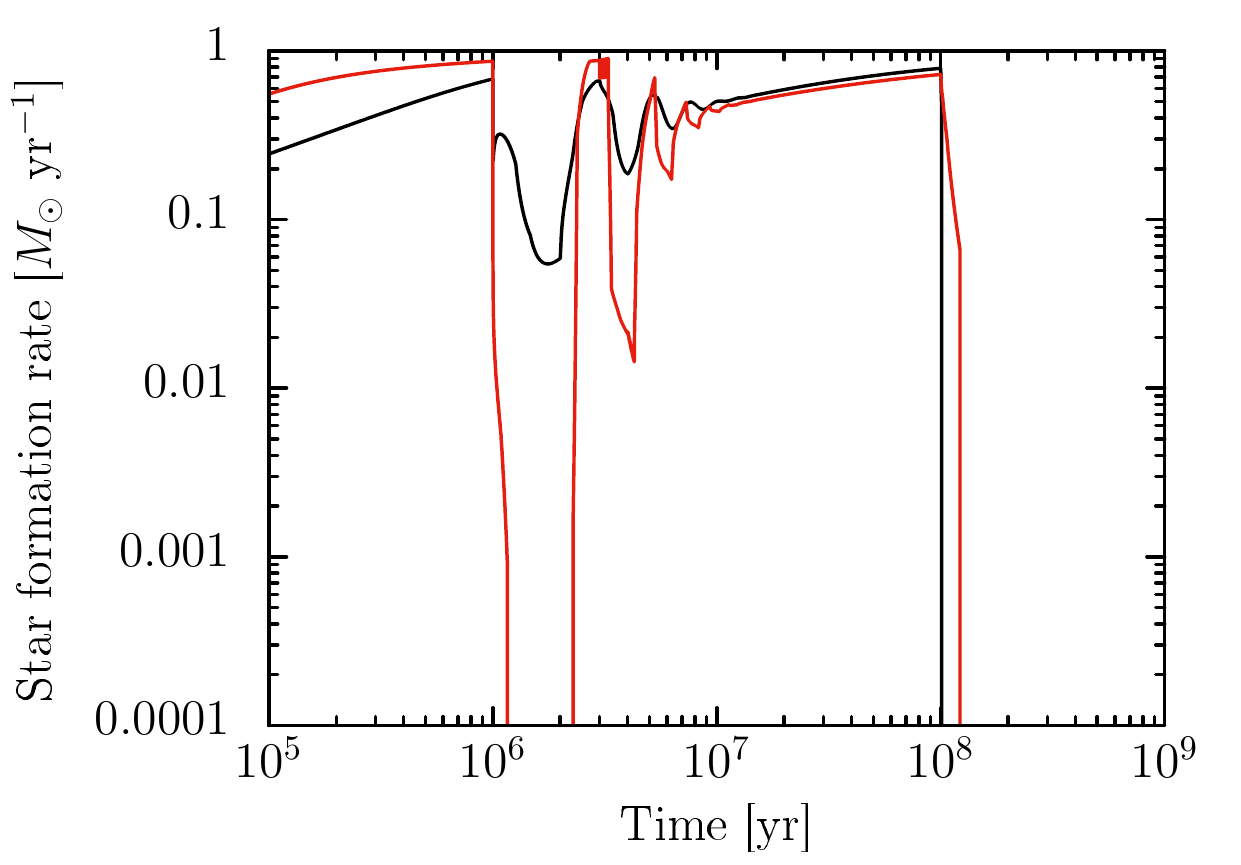}
         %\caption{b}
\end{minipage}
    \begin{minipage}{.5\linewidth}
    \hspace{4.5cm}
   \includegraphics[width=\columnwidth]{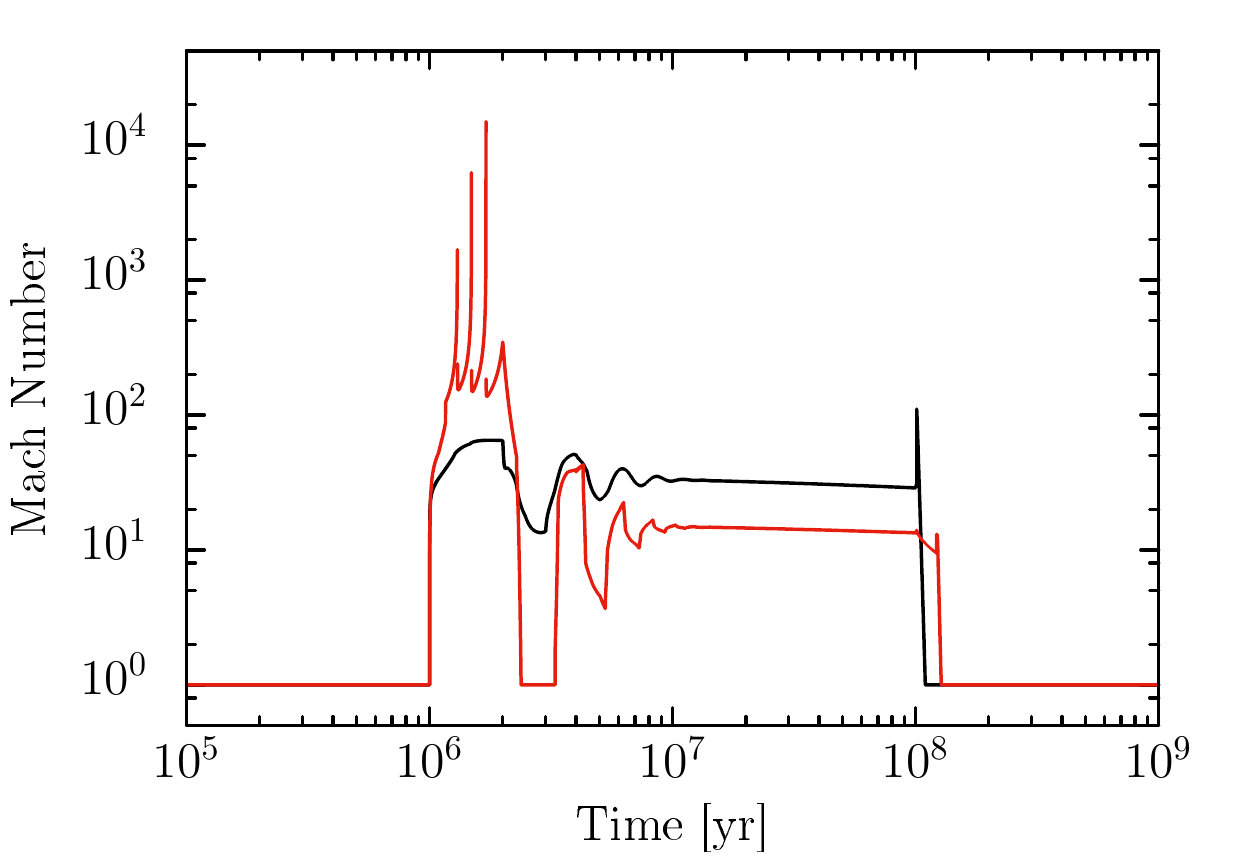}
         %\caption{c}
  \end{minipage}
    \caption{Time evolution of  $\dot{M}_{BH}(t)$, $\dot{M_{*}}(t)$, $M(t)$
  associated to model EB10 (red line), and model KM05 (black line) for a gravitational field dominated by the black hole.}
    \label{fig:2}
\end{figure*}

Although both models KM05 and EB10 produce similar final masses,  model KM05 produces more stable accretion during the high accretion phase whereas  model EB10
produces over-accretion during that phase (see figure \ref{fig:2}). Stable accretion is reflected  as less sharp fluctuations on the  accretion rate and on the star
formation rate in  the period of $10^{6}$ to $10^{7}$ years.  This effect is also reflected in the turbulent Mach number, which is driven by the feedback from star formation. 
Compared to the EB10 model, the dependence on the turbulent velocity is however less severe, scaling as $M^{-0.32}$ rather than $M^{-1}$, thus reducing the impact of the
negative feedback and damping the fluctuations within the system. In addition, as already described by \citet{Wutschik:2013caa}, there are hints of anticyclic variations 
between the black hole accretion rate and the star formation rate, which may be interpreted as competition between gas consuming processes  in the CND.

\begin{table}[t!]
  \centering
  \caption{Dependence of final black hole masses on initial seed, mass supply rate and dominant gravitational field.}
  \label{tab:2}
 \vspace{0.3cm}
 % \begin{tabular}{lllll}
  %\begin{tabular}{ |p{1.5cm}|p{1cm}|p{1.5cm}|p{1.5cm}|p{1.5cm}|  }
   \begin{tabular} {  m{1.5cm}  m{1cm} m{1.5cm}  m{1.5cm} m{1.5cm}   } 
    \hline
    \multicolumn{5}{r}{Final black hole mass [$10^{7}$ $M_{\odot}$]}\\
    \cline{3-5}
	Supply rate [$M_{\odot} \rm{yr}^{-1}$] & Seed $[M_{\odot}]$ & BH field &Host field & Disk field \\
	\hline\noalign{\smallskip}
	%\multirow{2}{*}{Defenders} 
	1&	$10^{3}$ &	2.95&	1.82&	3.09\\ 
	&	$10^{6}$&	2.96&	1.86& 	3.13\\  \hline
       10&	$10^{3}$&	23.3& 77.5& 15.6 \\
       &	$10^{6}$&	23.4& 	77.8& 15.6 \\ \hline
       100&	$10^{3}$&	118&	319& 	733\\
       &	$10^{6}$ & 118&	 319& 733\\
	\hline
  \end{tabular}
\end{table}

We have also  investigated the  influence of the  black hole seed mass on the system's evolution. For this purpose, table \ref{tab:2}~compares the final black hole  masses 
at adopting different seeds and under different assumptions regarding the central source of gravity. In all cases, it is clearly visible that the final black hole mass has an 
extremely weak dependence on the mass of the initial seed, even when considering seed masses of $10^{6}$ $M_{\odot}$, which are at the upper limit of what is suggested in 
the literature. This has been reported already by \citet{Armijo:2010md}, and \citet{Wutschik:2013caa}, the last one adopts different assumptions on star formation. 
However, as it was shown in figure \ref{fig:1}, a large mass supply rate to the system indeed enhances the final black hole masses. For instance for a supply rate of 
100 $M_{\odot}$ $\rm{yr}^{-1}$, the final masses  are of the order of $10^{9}$ $M_{\odot}$ which is in agreement with the results given by \citet{Volonteri2015ApJ}. While for moderate
mass supply rates of 1$M_{\odot}$~yr$^{-1}$ and 10 $M_{\odot}$~yr$^{-1}$, the final black hole masses also just weakly depend on our assumptions concerning the  gravitational acceleration, they 
differ by roughly a factor of~3 when considering large supply rates of 100 $M_\odot$~yr$^{-1}$.

\subsection{Evolution of the Mach number}

\begin{figure*}
\begin{minipage}{.5\linewidth}
\centering
 \includegraphics[width=\columnwidth]{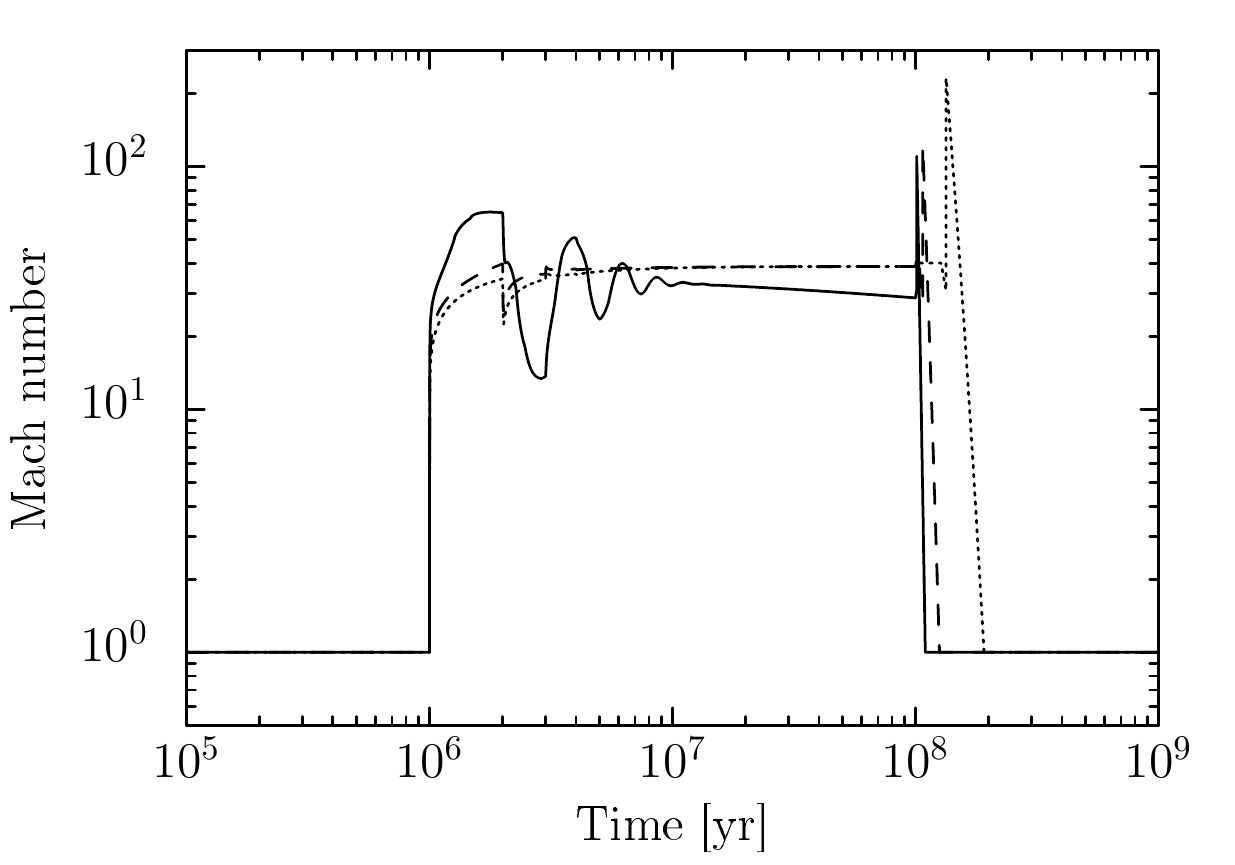}
  \caption{Mach number evolution with gravity dominated by the central black hole for different mass supply rates: 1 $M_{\odot}$ yr$^{-1}$ (line),~
  10 $M_{\odot}$ $\rm{yr}^{-1}$ (dashed line), 100 $M_{\odot}$ $\rm{yr}^{-1}$ (dotted line).}
      \label{fig:3}
\end{minipage}
\hspace*{0.2 cm}
\begin{minipage}{.5\linewidth}
\centering
\includegraphics[width=\columnwidth]{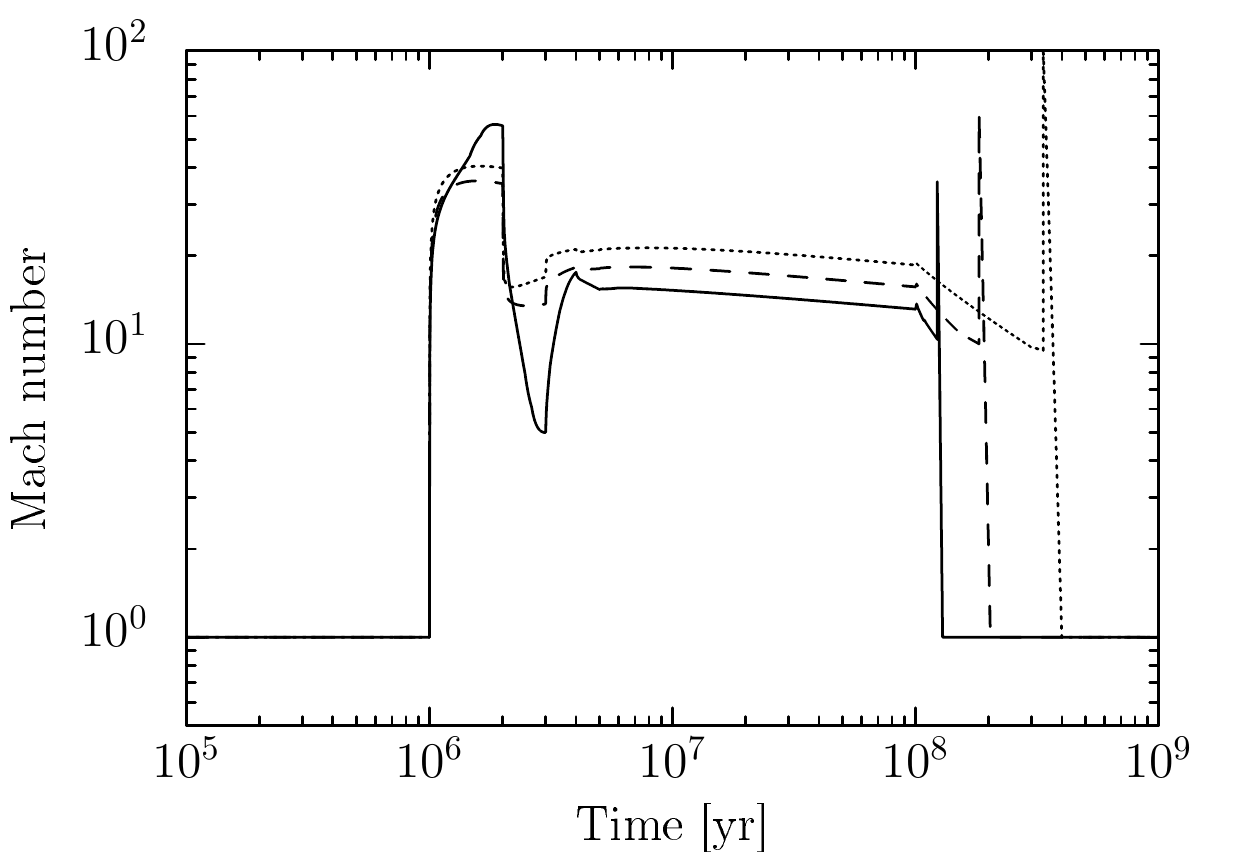}
  \caption{Mach number evolution with gravity dominated by the CND for different mass supply rates: 1 $M_{\odot}$ $\rm{yr}^{-1}$(line),
      10 $M_{\odot}$ $\rm{yr}^{-1}$(dashed line), 100~$M_{\odot}$ $\rm{yr}^{-1}$(dotted line).}
  \label{fig:4}
    \end{minipage}
   \begin{minipage}{.5\linewidth}
   \centering
\includegraphics[width=\columnwidth]{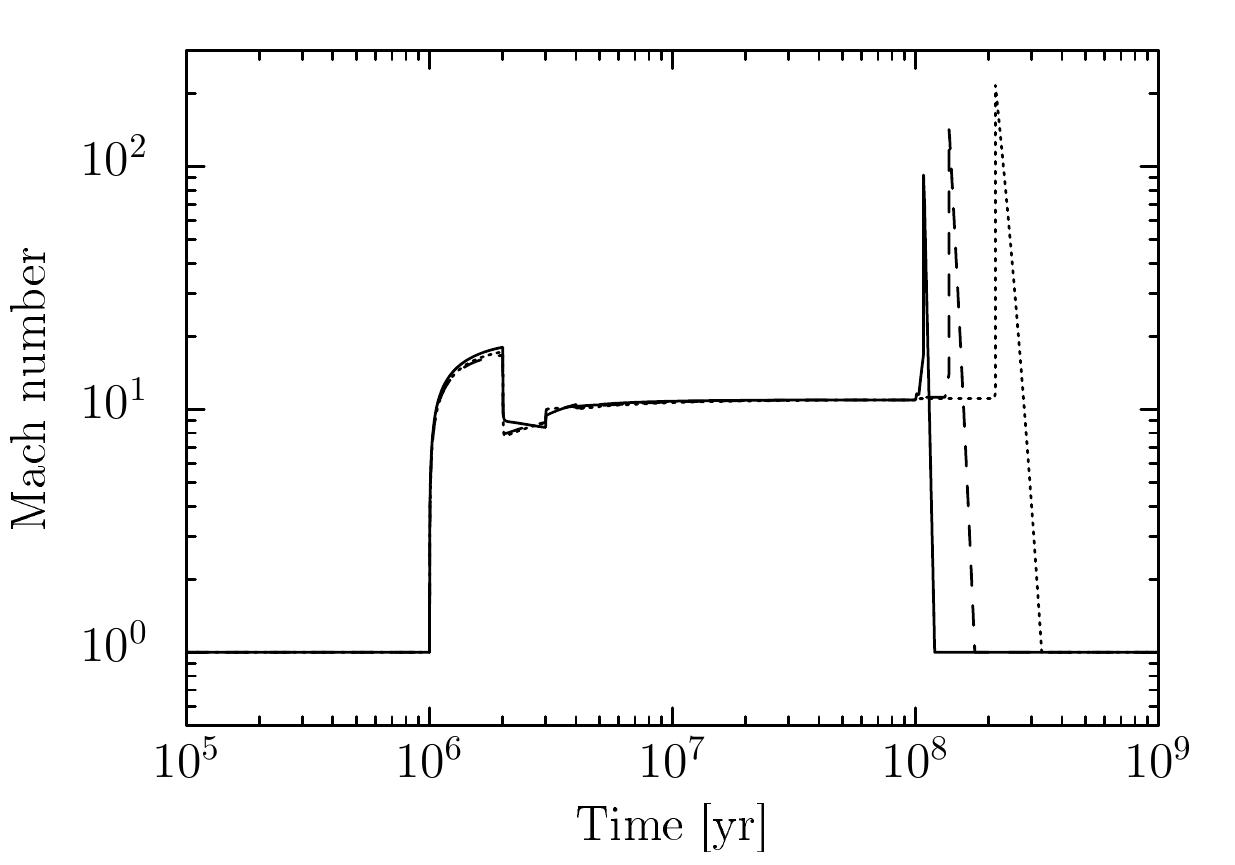}
  \caption{Mach number evolution with gravity dominated by the galactic disk for different mass supply rates: 1 $M_{\odot}$ $\rm{yr}^{-1}$(line),
      10 $M_{\odot}$ $\rm{yr}^{-1}$(dashed line), 100 $M_{\odot}$ $\rm{yr}^{-1}$(dotted line).}
        \label{fig:5}
\end{minipage}
\hspace*{0.2cm}
\begin{minipage}{.5\linewidth}
\centering
\includegraphics[width=\columnwidth]{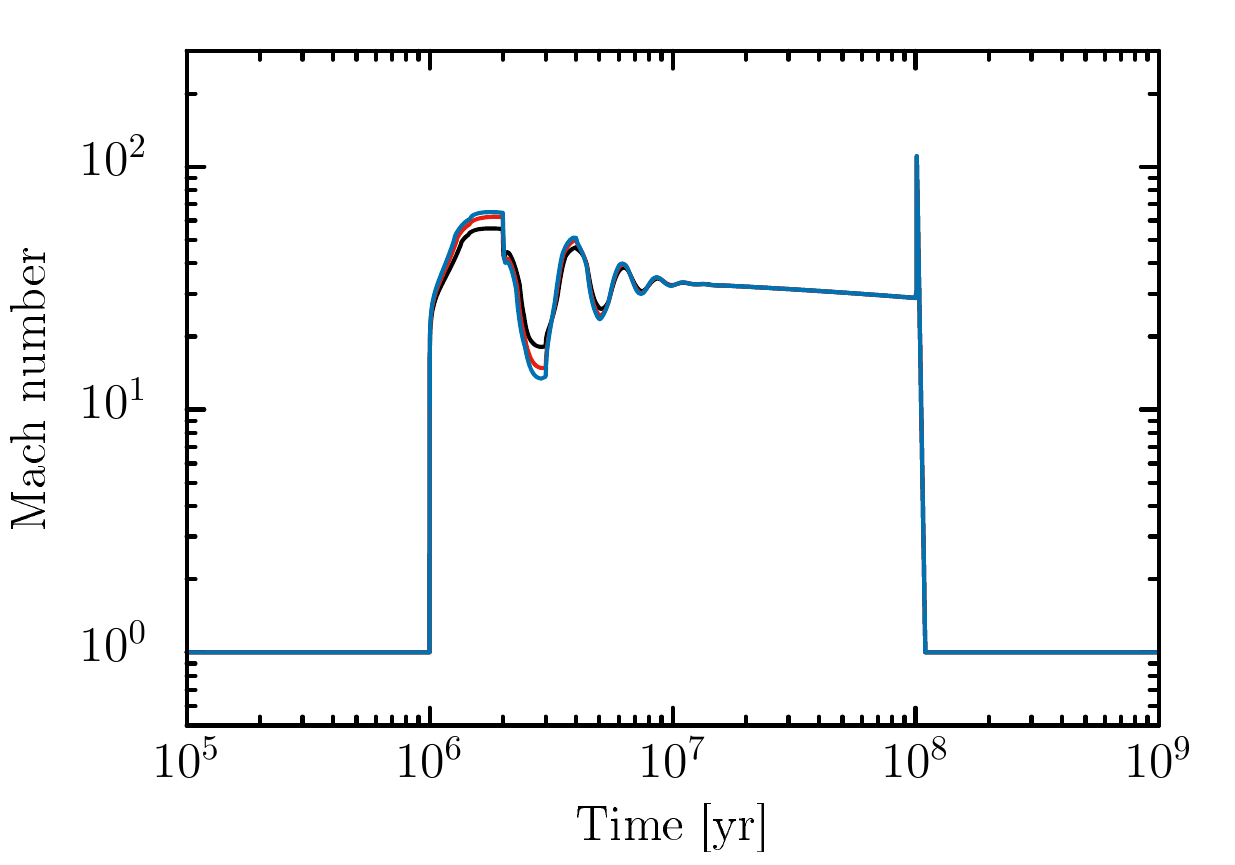}
  \caption{Mach number evolution with gravity dominated by the central black hole for different viscosity coefficients: $\alpha=1$ (blue line),
  $\alpha=0.5$ (red line), $\alpha=0.01$ (black line). Here the supply rate is  1$M_{\odot}$ $\rm{yr}^{-1}$.}
    \label{fig:6}
    \end{minipage}
    \end{figure*}

Figures \ref{fig:3}, \ref{fig:4}, and \ref{fig:5} show the  time evolution of the Mach number for different mass supply rates under different assumptions of the 
central gravitational field. We note in general that the turbulent velocity of the gas  becomes larger compared  to the local sound speed during  the high accretion 
phase (starting at $10^{6}$~years), which leads to large Mach numbers. When comparing the Mach number evolution in the different dominant gravitational scenarios, 
we notice that it is highest when the gravitational field is dominated by the central black hole, with $M_{\rm{max}} \approx 65$ at 
$\dot{M}_{sup}=$1~$M_{\odot}$ $\rm{yr}^{-1}$. In this scenario, the black hole accretes efficiently with $\dot{M}_{\rm{BH}} \approx$ 1$ M_{\odot}$ $\rm{yr}^{-1}$(see also Fig.~11). 

When the gas supply ceases after $10^{8}$ years, the  gas  accretion becomes less efficient, so the black hole accretes afterwards at very low rates 
(see figure \ref{fig:11}). At this timescale, a very pronounced spike appears on the Mach number evolution at all field dominations. The latter is a result of delayed 
supernova feedback at a time where the supply of new gas is inhibited, thus strengthening the effect of the feedback. Subsequently, on a timescale comparable 
to the lifetime of massive stars, the Mach number drops to a value of order unity, as turbulence is no longer driven by supernovae, but may result from hydrodynamical
instabilities. 
 
In the simulations presented here, the value of the viscosity coefficient $\alpha$ was of the order of unity in all our simulations. However, this quantity is still uncertain, and
it is worth mentioning that \citet{Kawakatu:2008rg}, and \citet{Wutschik:2013caa} have associated different 
values of $\alpha$ to the different accretion phases. In their assumptions, $\alpha$ has the order of unity in the turbulent supersonic 
phase  whereas in the subsonic phase $\alpha\approx0.01-0.5$ (due to magneto-rotational instabilities) where 
the turbulent velocity is comparable to the local sound speed. To study the dependence of our results on $\alpha$, we have run simulations to compare different values 
of $\alpha$ with $\alpha={0.01, 0.5, 1}$. Figure~\ref{fig:6} shows the time evolution of the Mach number for the different values of $\alpha$. It can be seen that small
values of $\alpha<1$  have a weak impact on the accretion phases, i.e. the gas velocity remains
comparable to the sound speed during the subsonic phase. Additionally, the Mach number evolution follows the same trend for all
values of $\alpha$ except that it reduces very slightly the Mach numbers in the period of 1 million, and 3 million years.
We find here that the time evolution of the BH accretion rate, star formation rate, final masses, and radius are 
quite independent of the parameter $\alpha$. Such independence on $\alpha$ shows that the accretion is dynamically regulated by the available mass reservoir, and 
therefore remains efficient as long as sufficient matter  will be available.

\subsection{Evolution of the CND radii}

\begin{figure*}
 \begin{minipage}{.5\linewidth}
  \centering
\includegraphics[width=\columnwidth]{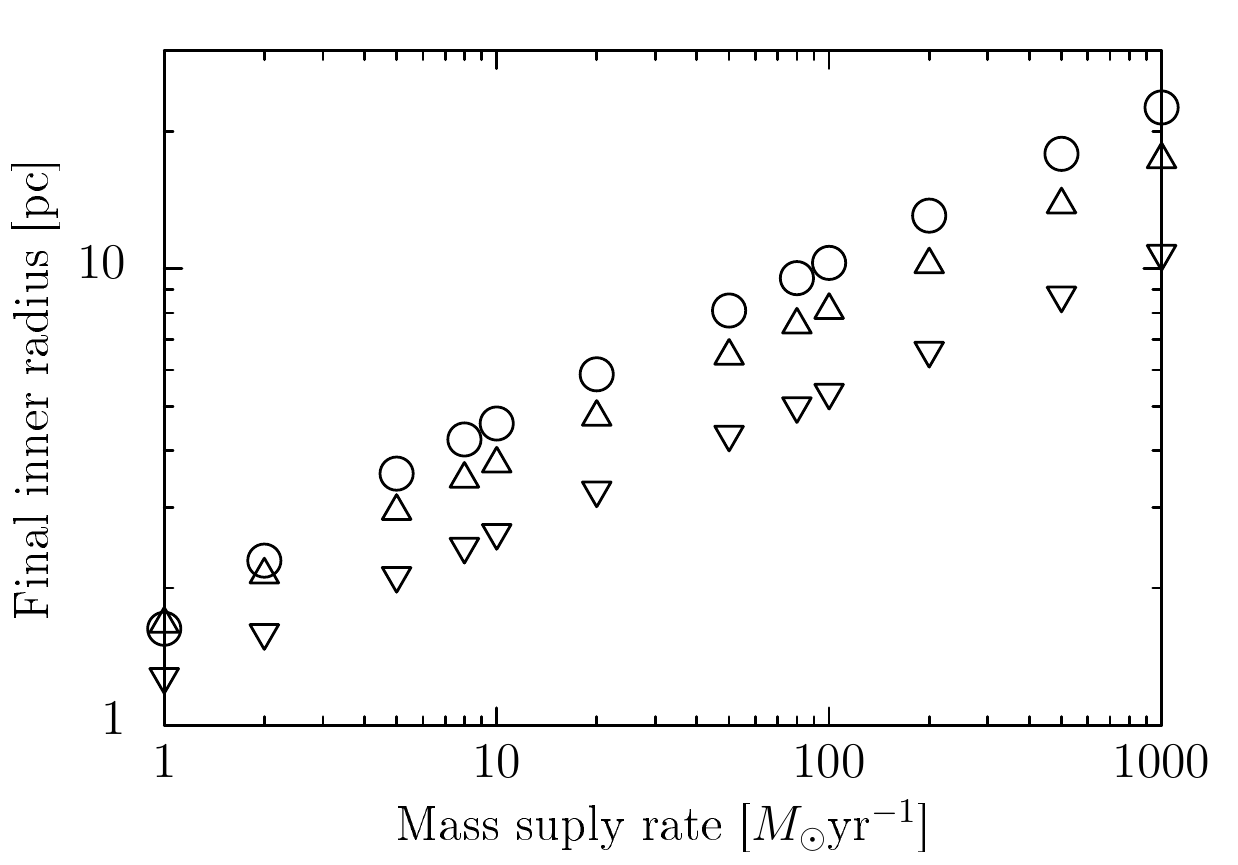}
  \caption{Disk inner radius vs. mass supply rate with gravity dominated by the central black hole (circles), with by the 
  CND (triangles), and by the host galaxy (inverted triangles).}
  \label{fig:7}
  \end{minipage}
   \hspace*{0.2cm}
 \begin{minipage}{.5\linewidth}
  \centering
\includegraphics[width=\columnwidth]{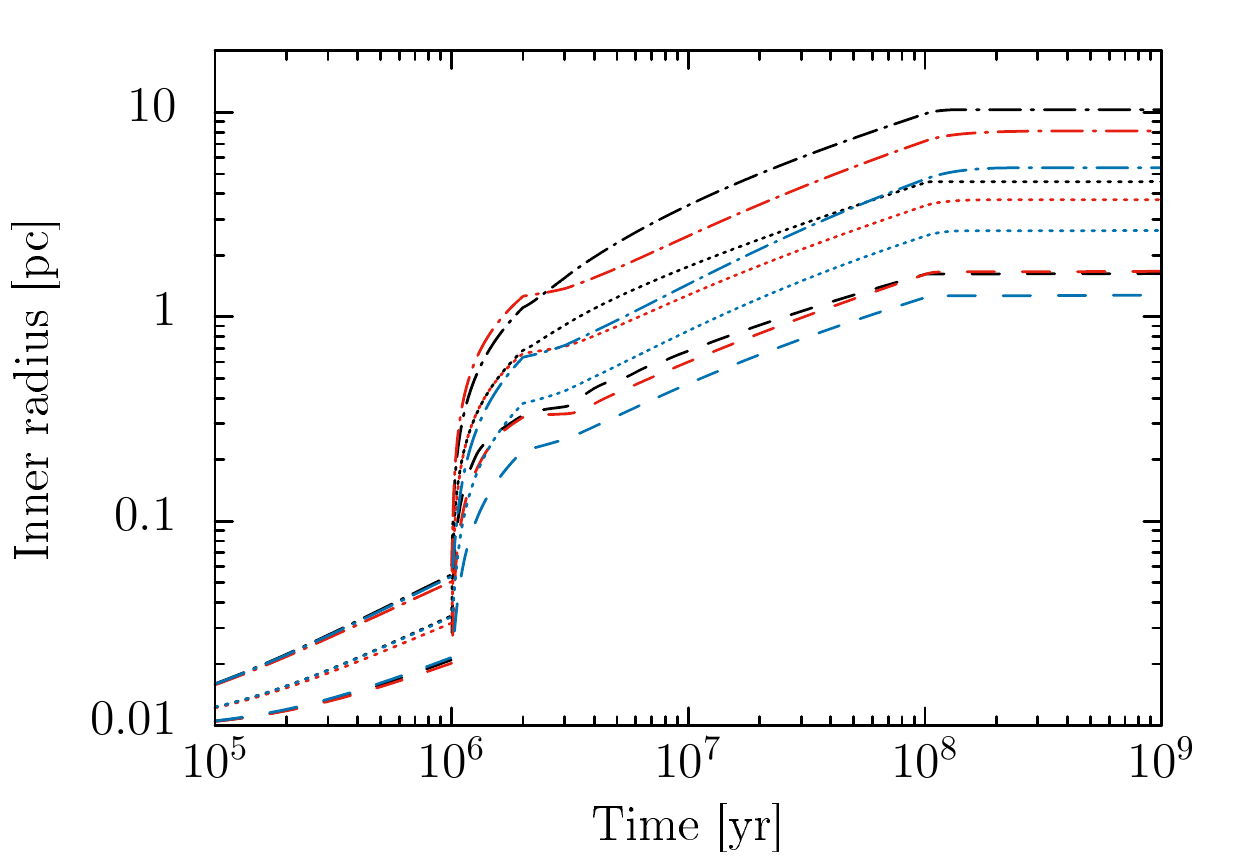}
  \caption{Time evolution of the disk inner radius for different mass supply rates: 1 $M_{\odot}$ $\rm{yr}^{-1}$ (dashed-line) , 10 $M_{\odot}$ $\rm{yr}^{-1}$ (dotted-line), 
  and  100~$M_{\odot}$~$\rm{yr}^{-1}$ (dashed-dotted line) with different assumptions on the central gravitational field: dominated by the central black hole (black color), 
  the CND (red color) and the  host galaxy (blue color).}
  \label{fig:8}
  \end{minipage}
   \hspace*{0.2cm}
 \begin{minipage}{.5\linewidth}
  \centering
\includegraphics[width=\columnwidth]{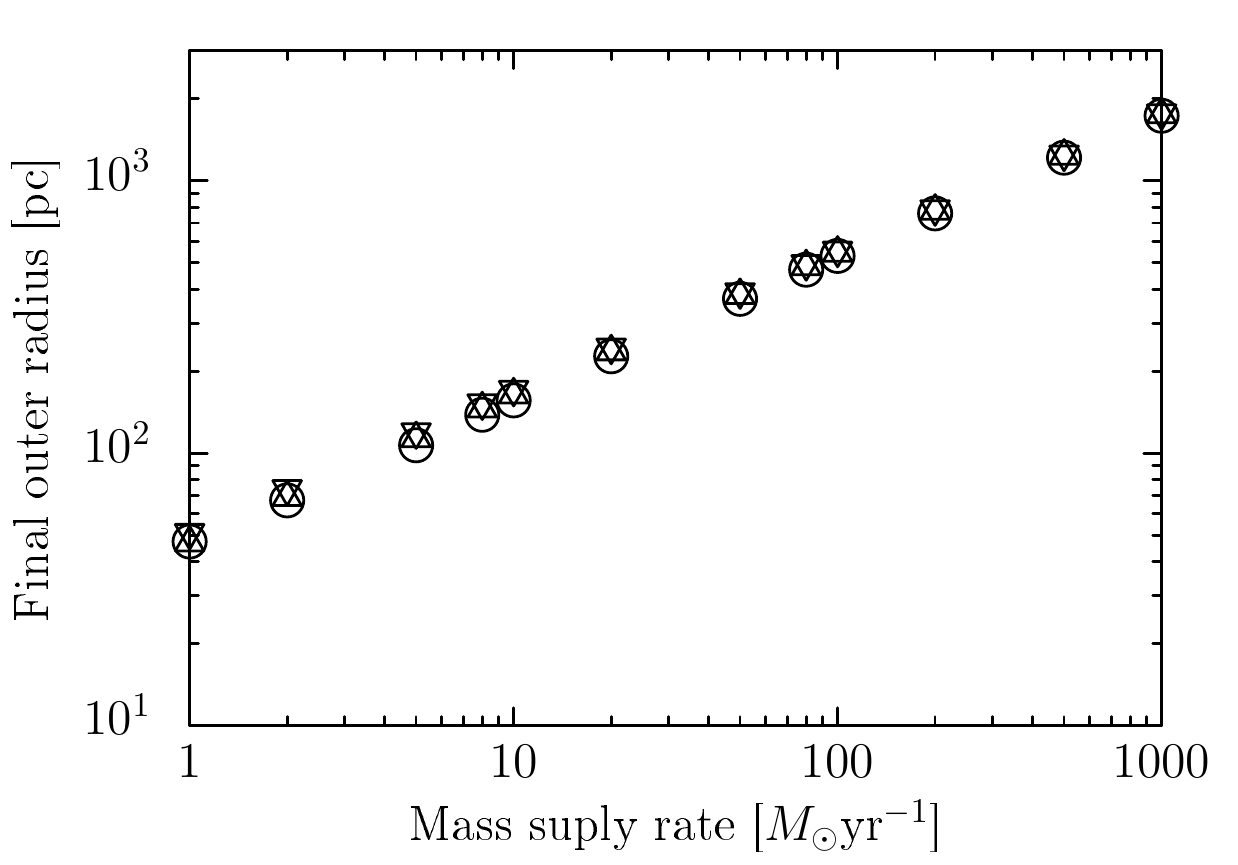}
  \caption{Disk outer radius vs. mass supply rate for gravity dominated by the central black hole (circles), by the CND (triangles) and by the host galaxy (inverted triangles).}
  \label{fig:9}
  \end{minipage}
   \hspace*{0.2cm}
 \begin{minipage}{.5\linewidth}
  \centering
\includegraphics[width=\columnwidth]{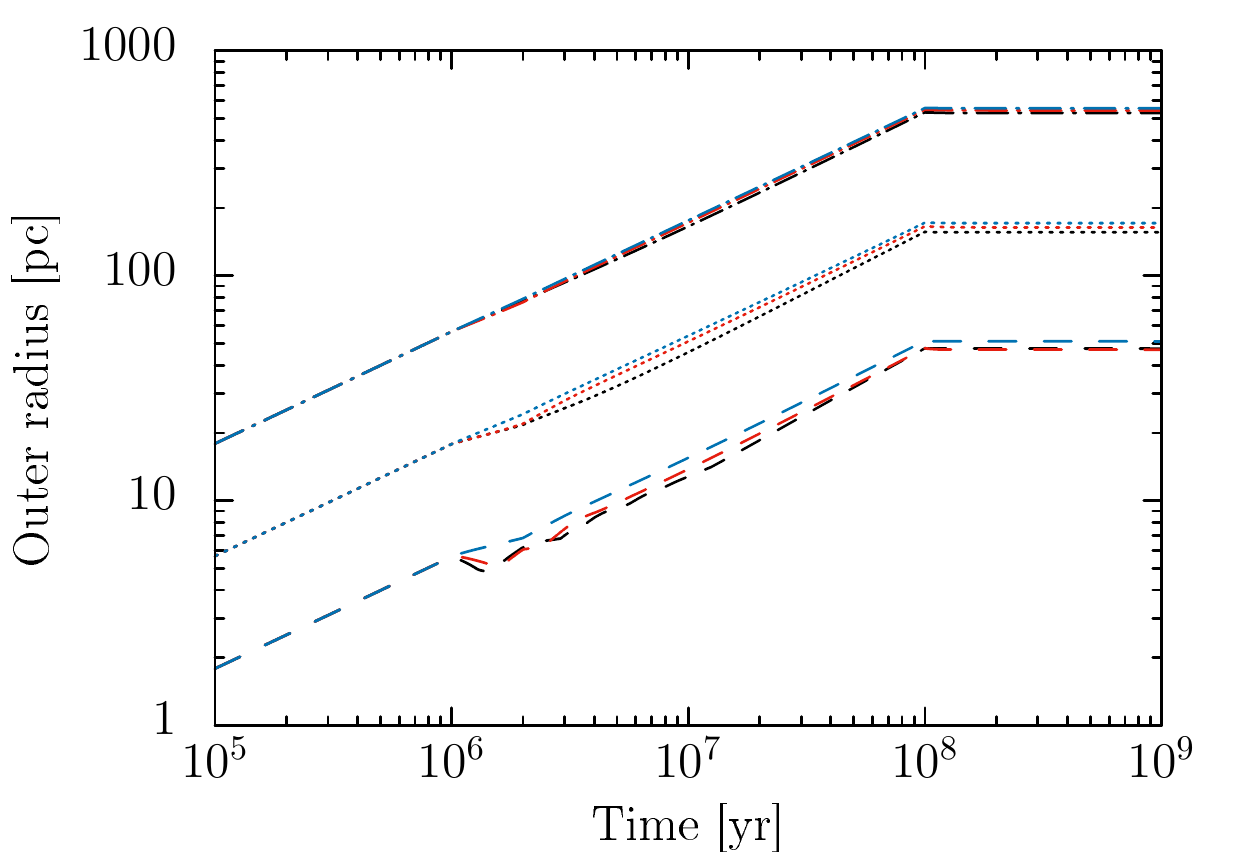}
  \caption{Time evolution of the disk outer  radius for different mass supply rates: 1 $M_{\odot}$ $\rm{yr}^{-1}$ (dashed-line) , 10 $M_{\odot}$ $\rm{yr}^{-1}$ (dotted-line) and  
  100~$M_{\odot}$~ $\rm{yr}^{-1}$ (dashed-dotted line) under different assumptions regarding the gravitational field, i.e. dominated by the central black hole (black color), by 
  the CND (red color) and by the  host galaxy (blue color).}
  \label{fig:10}
  \end{minipage}
  \end{figure*}

Returning to the evolution of the accretion disk itself, we have also studied the evolution of the final  inner and outer radius as a function of  time and  mass supply
rate (see figures \ref{fig:7} to \ref{fig:10}), exploring different assumptions regarding the gravitational field.  The explored  mass supply rates ranging  from 1 to 
1000 $M_{\odot}$ $\rm{yr}^{-1}$ allow the rapid growth of the disk radii  from parsec to kiloparsec scale. We find  no significant dependence of the  final values of the 
outer radius on the assumption regarding the gravitational field. So the results converge to similar values after the gas supply ceases at $10^{8}$ years. The final outer 
radius is comparable for instance with the size of some circumnuclear rings in galaxies like NGC~1097 with 1.5~kpc of radius studied
by \citet{Hsieh:2011ki}~and NGC 613 with a nuclear starburst of 700 pc wide found by \citet{falcon:2013}. The inner radius, on the other hand, always
remains at similar values,  with distances from the center spanning from 1 to approximately up to  30~pc.

\subsection{Evolution of the luminosities}

We also explore here the implications of our model on observable quantities, including the luminosity from the AGN ($L_{\rm{AGN}}$), the nuclear starburst ($L_{\rm{NSB}}$), the
stellar luminosity ($L_{\rm{stellar}}$) and for comparison the Eddington luminosity ($L_{\rm{edd}}$). The AGN luminosity is calculated based on two well-known cases: the 
standard thin disk  \citep{Shakura:1972te} which is applied for the low accretion phase and the 
slim disks \citep{Abramowicz:1988sp, Abramowicz:2004vi} for the high accretion phase. Similarly as in \citet{Kawakatu:2008rg}, equations (16) and (17) show
the dependence of the luminosities on the accretion rate. In the standard disk aproximation, $L_{AGN}$ is given  as
\begin{equation}
\label{eq:16}
\rm{ L_{AGN}} =\left(  \frac{\dot{m}(t)}{10}\right) \rm{L_{edd}(t)}, \enspace  {\rm \dot{m}(t)< 20,}
\end{equation}
and in the slim disk approximation $L_{AGN}$ is given  as
\begin{equation}
 \label{eq:17}
 \rm{L_{AGN}} = 2  \left(  1+ln \frac{\dot{m}(t)}{20} \right) \rm{L_{edd}(t)},  \enspace {\rm \dot{m}(t)\geq20},
\end{equation}
where $\rm{{\dot{m}}=\frac{\dot{M}_{BH}}{\dot{M}_{edd}}}$, and  $\rm{\dot{M}_{edd}=\dot{M}_{crit}=\rm{L_{edd}}/c^{2}}$  \citep{Watarai2000}. The Eddington 
luminosity $\rm{L_{edd}}$ is given by the following expression:
\begin{equation}
\label{eq:18}
 \rm{L_{edd}}= 1.25 \times 10^{38} \frac{M_{BH}}{M_{\odot}} [erg/s].
\end{equation}
For calculating the nuclear  starburst luminosity ($\rm{L_{NSB}}$), we adopt the formula given by   
\citet{Kawakatu:2008rg}, with $\epsilon=0.007$:
\begin{equation}
\label{eq:19}
 \rm{L_{NSB}(t)}=0.14 \epsilon \dot{M}_{*}(t) c^{2}.
\end{equation}

Since our model includes phases without active star formation, we estimate the total stellar luminosity by adopting the mass-to-light ratio ($M/L$)
 from the total stellar mass.  Typical values range from 2 to 10 solar masses per solar luminosity \citep{Dickel:1978di, Faber:1979fa}. 
For our calculations we adopt a mass-to-light-ratio of 6 $M_{\odot}/L_{\odot}$. 

\begin{figure*}
 \begin{minipage}{.5\linewidth}
 \centering
 \includegraphics[width=\columnwidth]{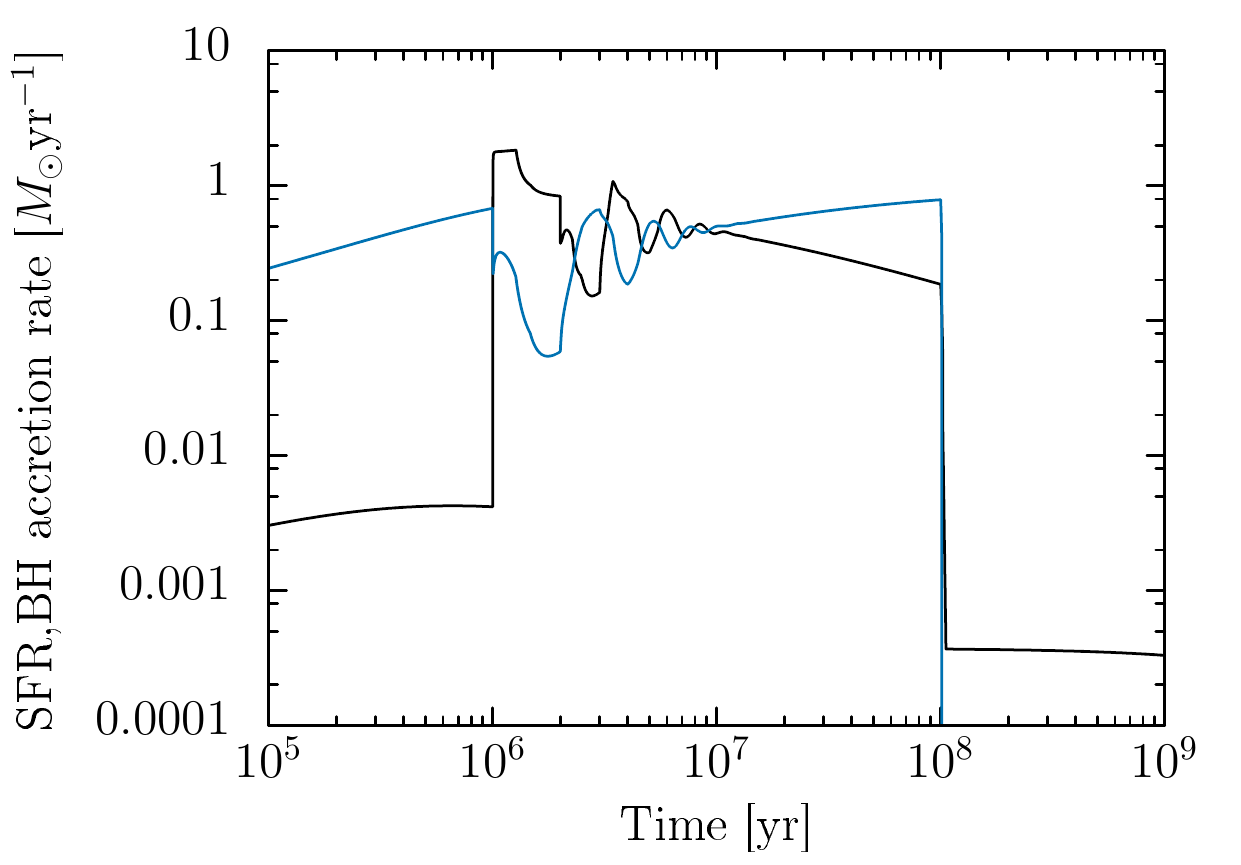}
  \caption{Time evolution of the black  hole (BH) mass accretion rate (black line) and star formation rate (SFR) (blue line) with gravity dominated by the central 
  black hole, assuming a mass supply rate of 1~$M_{\odot}$~$\rm{yr}^{-1}$.}
  \label{fig:11}
  \end{minipage}
  \hspace*{0.2cm}
   \begin{minipage}{.5\linewidth}
  \centering
\includegraphics[width=\columnwidth]{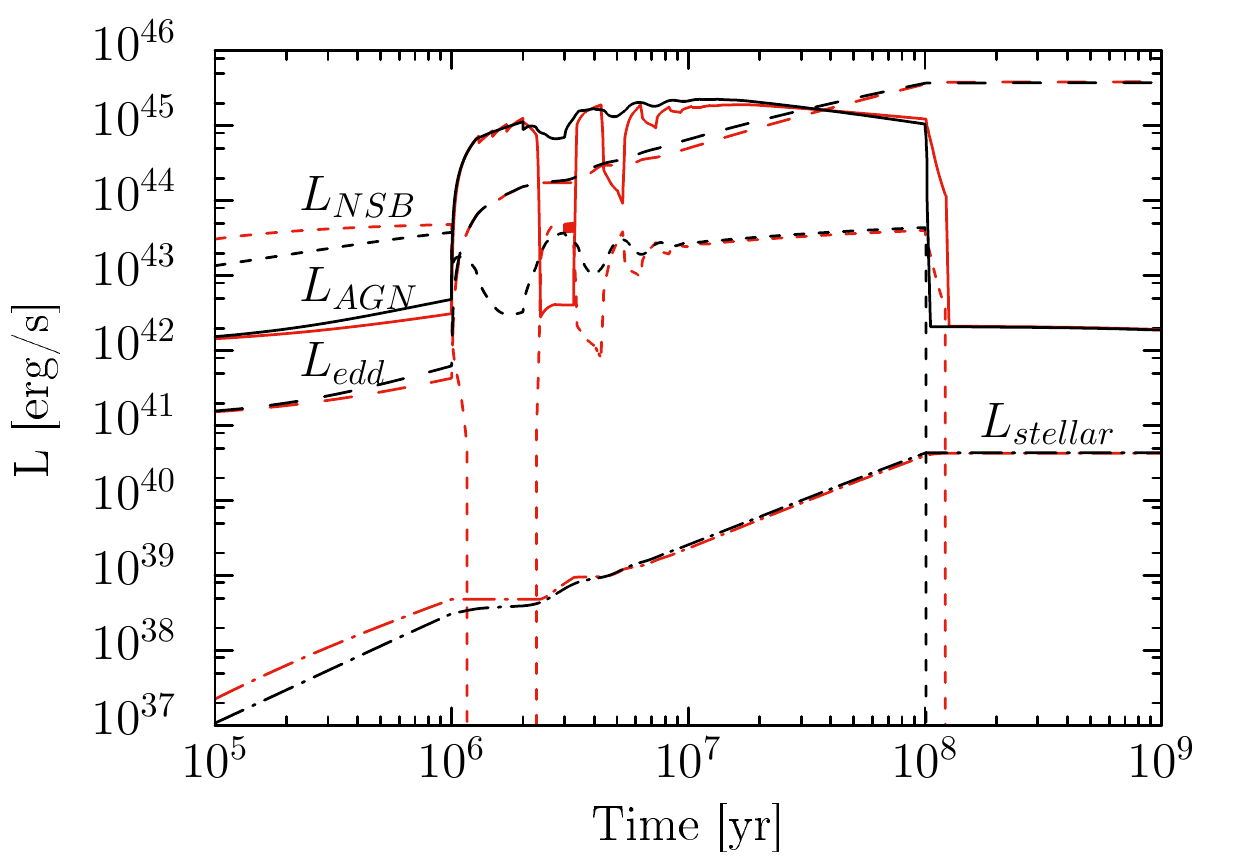}
  \caption{Comparison of the time evolution of the luminosities bewteen model KM05  (black curves) and model EB10  (red curves). The KM05 model is applied
  with gravity dominated by the central black hole, assuming a mass supply rate 
  of 1~$M_{\odot}$~$\rm{yr}^{-1}$. $\rm{L_{edd}}$   is the Eddington luminosity,  $\rm{L_{NSB}}$ the nuclear starburst luminosity, $\rm{L_{AGN}}$ is the AGN 
  luminosity and $\rm{L_{stellar}}$ is the total stellar luminosity.}
  \label{fig:12}
\end{minipage}
  \end{figure*}

The curves shown in figure 12  are the results of our simulations for an initial black hole seed mass of $10^{3}$ $M_{\odot}$,  with a mass supply rate 
of 1 $ M_{\odot}$ $\rm{yr}^{-1}$. The rest of the input parameters are listed in table \ref{tab:1}. Thus figure \ref{fig:12} compares the time evolution of the 
AGN luminosity, the Eddington luminosity, the  nuclear starburst luminosity and the total stellar mass
using model KM05 and model EB10. The AGN luminosity is calculated assuming gravity to be dominated by the central black hole. 
We note that in both models the time evolution of the Eddington luminosities are similar and by comparing them with AGN luminosites, we identify different 
timescales of the accretion phase: the high accretion phase ($10^{6}-10^{8}$~years) and the low accretion phase ($>10^{8}$ years). We further note that the
luminosities should more substantial fluctuations in the model EB10 compared to KM05, which is consistent with the results obtained for the star formation and 
accretion rates.

These results show that our  model predicts upper limits of the luminosities: the highest AGN  luminosity ($\sim 10^{45}$ erg/s) and the lowest AGN 
 luminosity ($\sim 10^{42}$~erg/s) as well as the highest nuclear starburst luminosity ($\sim 10^{43}$ erg/s), which are
 in agreement with \citet{Kawakatu:2008rg}. Furthermore, we note that our model predicts super-Eddington luminosities for the period between $10^{6}$ years 
 and $2\times 10^{7}$ years. Sub-Eddington luminosities occur when the  AGN luminosity starts to decrease in the period of  $2\times 10^{7}$ years to $10^{8}$ years, 
 then the AGN luminosity drops drastically to lower values after~$10^{8}$~years. 
 
 In general, we find no significant differences in the time evolution of the total stellar  luminosities and Eddington luminosites when models KM05 and EB10 are applied. 
 However,  the time evolution of the AGN and  nuclear starburst luminosities in both models differs during the high accretion phase. As we have described 
 in section 3.1 model EB10  produces over-accretion during the high accretion phase so in consequence these effect also produces the variabilites on the luminosities
 displayed in figure 12.

We want to emphasize that the typical timescale for activity in AGN depends on external parameters, and for instance, the 
external mass supply could last for a shorter time or longer time than $10^{8}$ years. This would depend  to some extend on the type of the AGN, the redshift and also the 
environment. For instance for local AGNs, it may be possible that the accretion phases are typically shorter, while there might
be rather long and extended periods of accretion necessary to explain high redshift quasars. 

\section{Application of the model to active galaxies and starburst rings}

\subsection{High redshift quasars}

 \cite{Mortlock:2011lo} discovered the most distant luminous quasar ULAS J1120+0641, which resides at~$z~=7.085$ ($7.6\times~10^{8}$~years after the big bang). The estimated
 mass of the supermassive black hole hosted by that quasar is~2~billion~solar~masses which generates a very large luminosity of~$2.4~\times~10^{47}$~erg/s.~Another distant
 quasar at $z=6.42$, J1148+5251 was discovered by \citet{Willot2003ApJ}, with a supermassive black hole mass  of $4.9\times~10^{9}$ $M_{\odot}$ and a luminosity 
 of $5.2\times 10^{47}$ $M_{\odot}$ \citep{Pezzulli2016MNRAS} (with references therein). We adapt here our model to  reproduce the observed AGN properties of these 
 two high redshift quasars. First,  we extended the time of mass supply to~$8\times10^{8}$~years, and run simulations for mass 
 supply rates of 30 $M_{\odot}$ $\rm{yr}^{-1}$ and 100 $M_{\odot}$ $\rm{yr}^{-1}$. The rest of the parameters corresponds to the values given in Table~1.

\begin{figure*}
 \begin{minipage}{.5\linewidth}
  \centering
\includegraphics[width=\columnwidth]{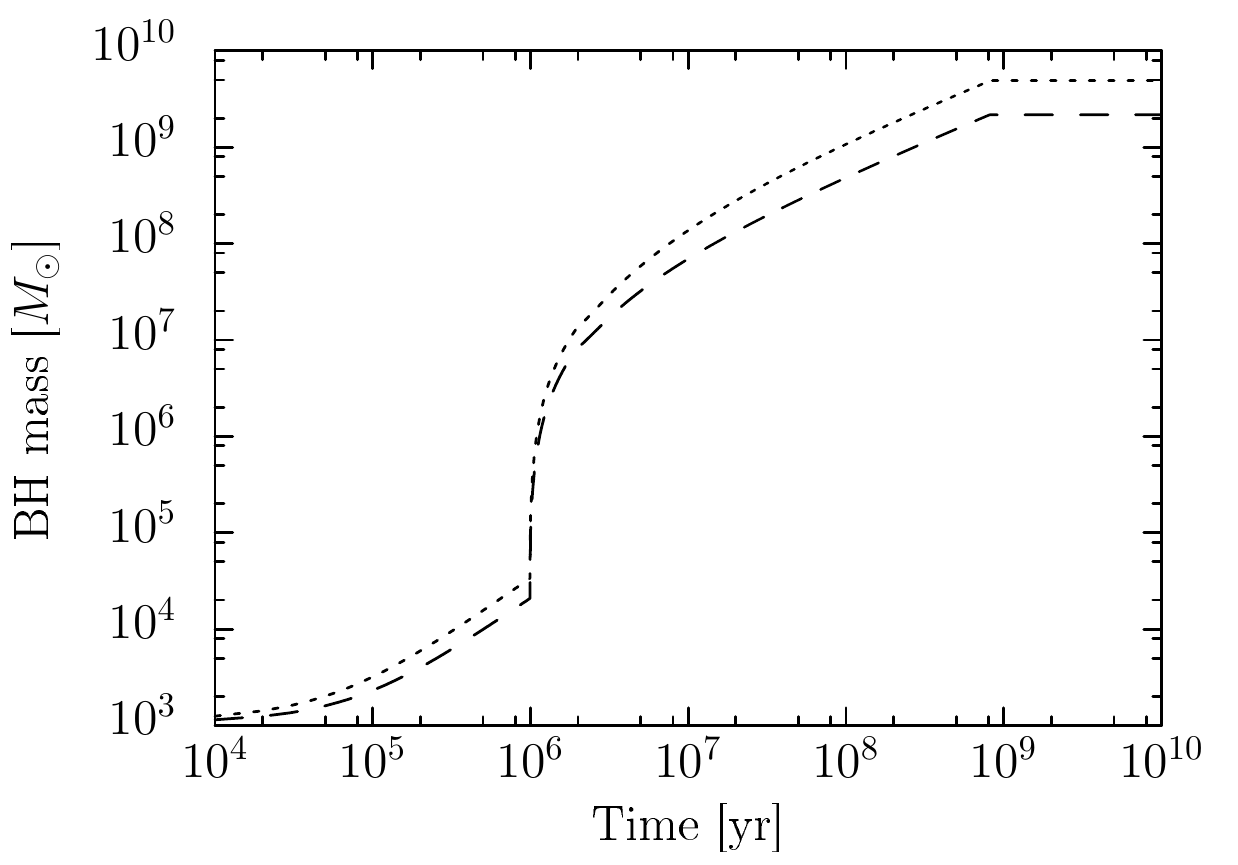}
 \caption{Time evolution of the black hole (BH) masses for different supply rates: 30 $M_{\odot}$ $\rm{yr^{-1}}$ (dashed line) and 100 $M_{\odot}$ $\rm{yr^{-1}}$(dotted line).
 . When the gas supply  ceases at  $8\times10^{8}$ years the final masses are reached. }
   \label{fig:13}
 \end{minipage}
 \hspace*{0.2cm}
  \begin{minipage}{.5\linewidth}
  \centering
\includegraphics[width=\columnwidth]{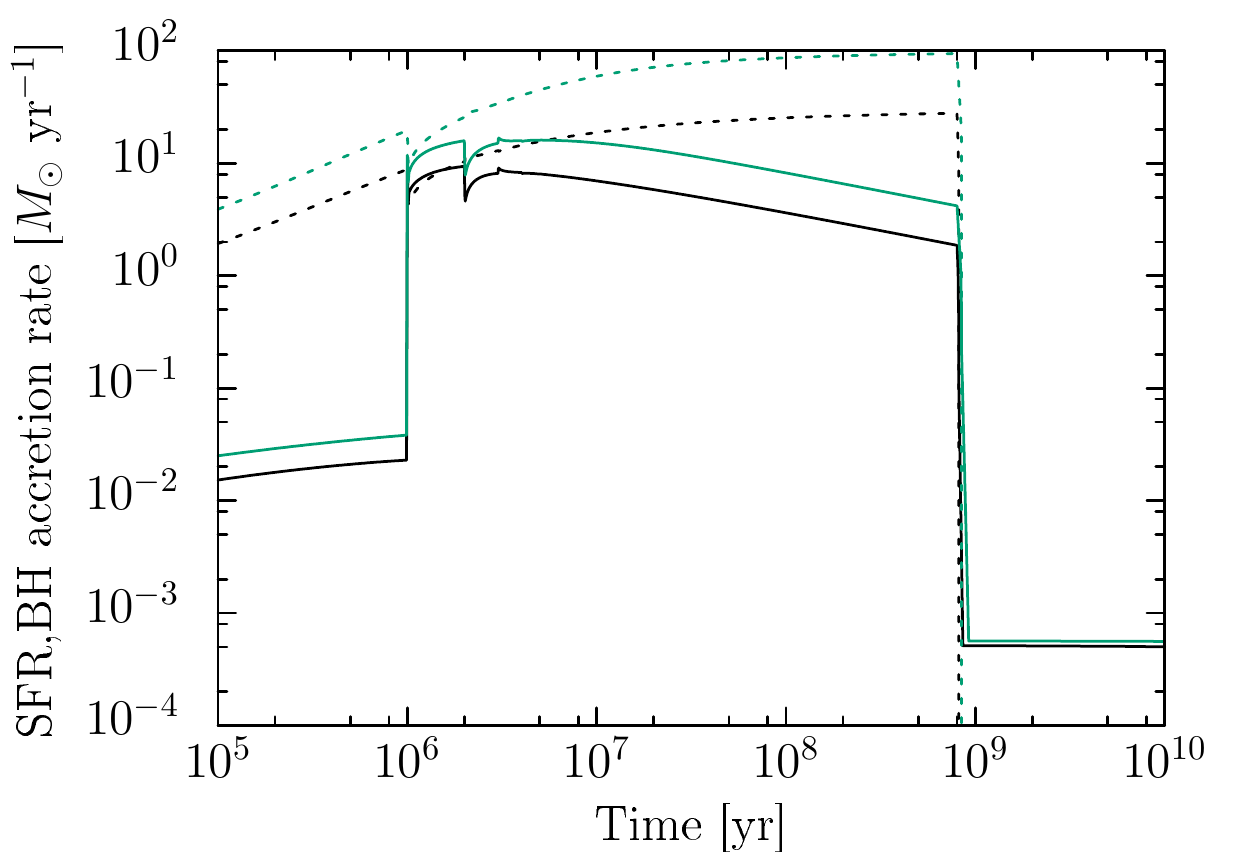}
  \caption{Black hole (BH) accretion rate (line) and star formation rate (SFR) (dotted-line) evolution for different supply rates.  At $8\times10^{8}$ years, supply 
  rates of 30 $M_{\odot}$ $\rm{yr}^{-1}$(black curves)  and 100 $M_{\odot}$ $\rm{yr}^{-1}$ (green curves)  produces  final black hole masses $2.2\times 10^9$ $M_{\odot}$ and
  $4.9\times 10^9$ $M_{\odot}$ respectively.}
    \label{fig:14}
\end{minipage}
\begin{minipage}{.5\linewidth}
  \includegraphics[width=\columnwidth]{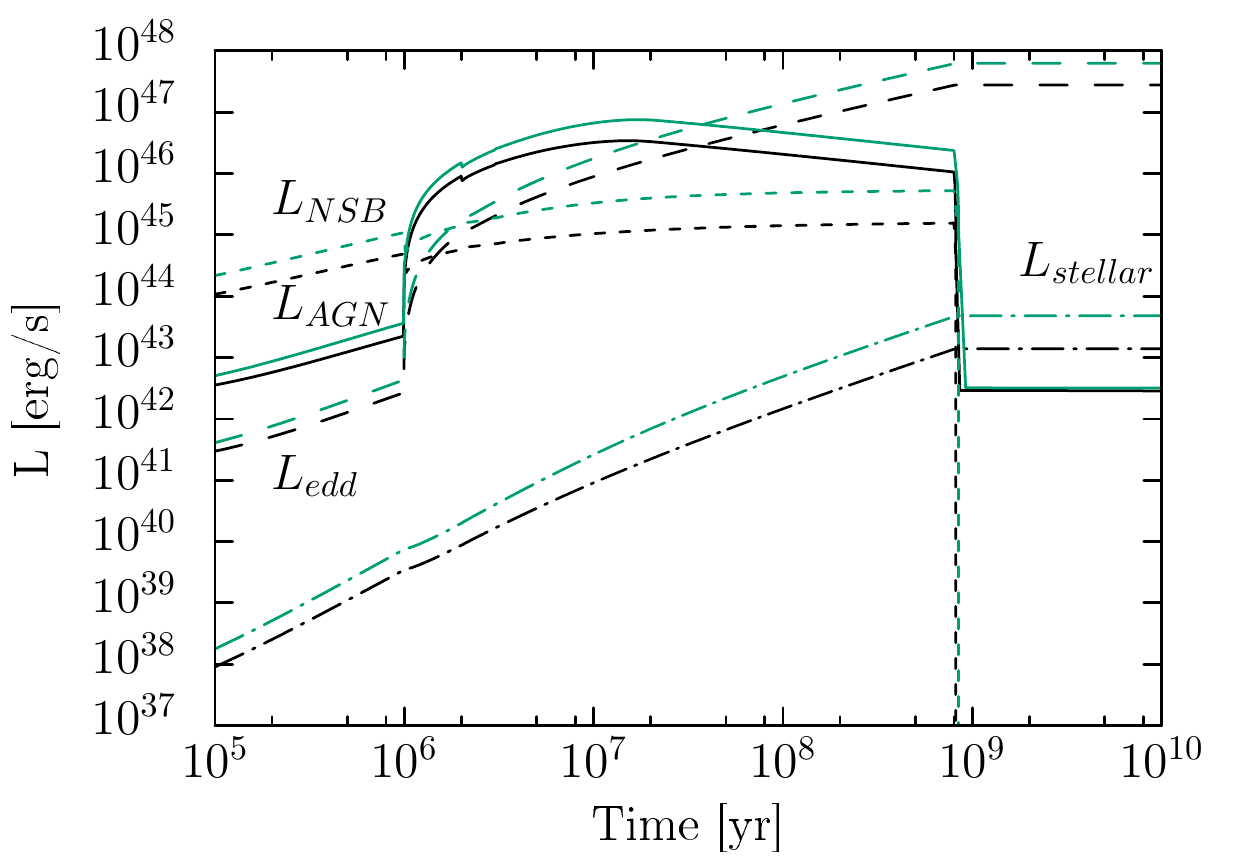}
      %\caption{a}
\end{minipage}
\begin{minipage}{.5\linewidth}
 \includegraphics[width=\columnwidth]{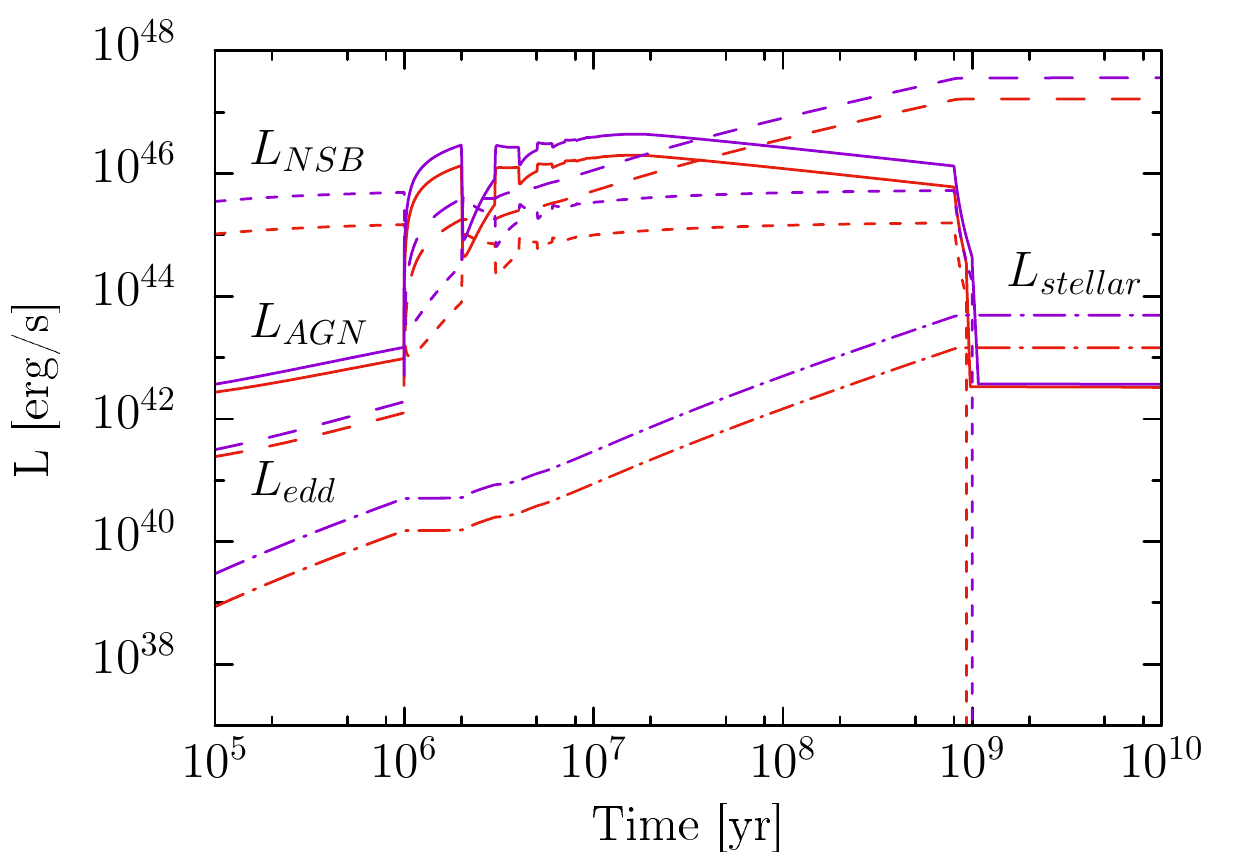}
         %\caption{b}
\end{minipage}
    \caption{Comparison of time evolution of the luminosities with model KM05 and model EB10.
    Left: Model KM05  for different supply rates: 30~$M_{\odot} $ $\rm{yr}^{-1}$ (black curves)  and 100 $M_{\odot}$ $\rm{yr}^{-1}$ (green curves).
    Right: Model EB10  for different supply rates: 30~$M_{\odot} $ $\rm{yr}^{-1}$ (red curves)  and 100 $M_{\odot}$ $\rm{yr}^{-1}$ (purple curves).}
    \label{fig:15}
\end{figure*}

As a result, our model  produces final black hole masses of  $2.2\times10^{9}$ $M_{\odot}$ and $4.91\times10^{9}$ $M_{\odot}$ when the gas supply ceases after
$8\times10^{8}$ years (see figure \ref{fig:13}). Such large mass supply rates allow efficient accretion up to 10~$M_{\odot}$ $\rm{yr}^{-1}$ (see figure \ref{fig:14}). This 
large accretion rate lasts only for a  short time in the high accretion phase and then produces an upper limit in the  luminosity  of almost  $\sim10^{47}$ erg/s at 
roughly $10^{7}$ years (see left graph of figure \ref{fig:15}). Shortly after the maximum luminosity is reached, the quasar emits continuous  radiation at a 
rate of $\sim10^{46}$ erg/s until the gas supply ceases. 

We find again that our model predicts super-Eddington and sub-Eddington luminosities even when the time supply is extended up to one billion years. These results are displayed
in figure 15, where we compare the time evolution of the luminosities using  model KM05 and model EB10 in the same fashion as in figure 12. 
We note that the application of model EB10 for  large accretion rates (>1 $M_{\odot}$ $\rm{yr^{-1}}$) reduces the over-accretion, but it still produces variations of the
AGN and nuclear starburst luminosities during the high accretion phase.

Figure \ref{fig:14} shows that the star formation rate is the preferred mode of gas consumption in the high accretion phase. This effect can be seen 
in the period of $10^{6}$ years (SN feedback timescale) and $10^{9}$ years, where  the evolution of the star formation rate comprises larger and enhanced rates  than 
the BH accretion rates. This behaviour can be attributed  to the  very large mass supply rate injected from the host galaxy to the outer disk of the CND, where the stars 
are forming. It is important to note that the injection of a large mass supply rate produces little gas competition between the BH accretion rate and  the SFR, whereas 
for low supply rates, as it was shown in the previous section, a  high  competition for gas consumption and strong fluctuations develop as displayed in  figure \ref{fig:11}. 
Hence, our model predicts an upper limit of the SFR value, which is close to $100$~$M_{\odot}$ $\rm{yr}^{-1}$ (see figure \ref{fig:14}). The observed values of the SFR in the 
quasars at $z=7.085$ and $z=6.42$ are 160 - 440 $M_{\odot}$ $\rm{yr}^{-1}$ and 2000 $M_{\odot}$ $\rm{yr}^{-1}$ respectively, which were estimated both from far-infrared (FRI) 
luminosities reported by \citet{Venemans2012ApJ} and \citet{Pezzulli2016MNRAS} (with references therein). While the comparison of 
the  SFR values predicted by our model with the observed results shows an apparent discrepancy, one should note that the quoted values of the SFR refer to the entire host
galaxy, while our model only follows the evolution in the CND. Higher-resolution observations may however help to constrain the SFR in the central region of the galaxy in the 
future.

\subsection{The prominent starburst ring in  NGC 6951}

NGC 6951 is a Seyfert 2 galaxy ($z=0.004750$) with a  barred spiral morphology. It exhibits a characteristic starburst ring with radius of 
580~pc  \citep{vanderLaan:2013qy} but other studies have claimed a radius of 480 pc \citep{Sani:2012sani}. Van der Laan et al. (2013) has determined the stellar content on 
the circumnuclear rings that  spans from young to old stars. The stellar clusters
along the ring contain  predominantly massive stars with intermediate ages. The total stellar mass estimated  in the ring is~$6\times10^{7}$~$M_{\odot}$ with an estimated 
age  between 1 Gyr and 1.5 Gyr. No other determination of the age of the ring has been published so far.

The mass of the central black hole ranges between~$5.5\times~10^{6}$~$M_{\odot}$ at $81^{o}$ and $1.3\times 10^{7}$ $M_{\odot}$ at $33^{o}$ 
\citep{Beifiori:2012}. As the angle under which we see the system is still uncertain,  the unknown angle leads to an uncertainty in the dynamical quantities, in spite of 
the position of the AGN nucleus and its $H_{\beta}$ luminosities having been accurately measured by~\citet{Perez:1999je}.

%addtional text
The presence of a bar in the galaxy can potentially affect the formation of a circumnuclear disk and partly influence its evolution. As previously shown by 
\citet{Athanassoula1992MNRAS328A} and \citet{Regan2003}, the existence of a circumnuclear ring inside a large-scale bar depends on the orbit families that make up the 
circumnuclear region. While the x1 orbits are elongated along the asymmetric potential and make up the stellar bar, the x2 orbits are oriented perpendicular to the bar, and 
gas-orbits turn slowly from one into the other within a bar potential. While we cannot exclude that this has an impact on the disk, our previous analysis in section 3 has 
shown that both the stellar mass and the black hole mass are relatively insensitive to the details of the potential. With these caveats in mind, we therefore tentatively 
compare our model to NGC 6951, knowing that a description of the full dynamics should also consider the bar in more detail.

We  apply and adapt here our model to reproduce the properties of the ring in NGC 6951 related to the stellar mass, the radius of the ring, the evolution of the star 
formation rate  as well as the central  SMBH mass. In our simulations, we study the evolution of the CND with an initial black hole seed of~$10^{3}$~$M_{\odot}$. The time 
for mass supply from the host galaxy is extended up to a billion years. We have adopted a low surface density of the  host galaxy as 200 $M_{\odot}$ $\rm{pc}^{-2}$ and a low constant mass
supply rate of 0.2 $M_{\odot}$ $\rm{yr}^{-1}$. In addition, we have run the simulations for different timescales associated to  supernovae explosions (SN) 
produced by nuclear starbursts (massive stars). Here we explore the effect of varying the average lifetime of the massive stars, which may be considered as revealing the effect
of different stellar populations ending their life at different times. In particular, we adopt possible lifetimes of  $5\times 10^{6}$ years (21 $M_{\odot}$), $10^{7}$ years (15 $M_{\odot}$)
and~$5\times~10^{7}$~years (8 $M_{\odot}$). With 
these assumptions, our model  is able to reach good agreement with the observed properties of the circumnuclear ring in NGC 6951. Table \ref{tab:3} shows the final 
results of the BH mass, mass of the ring and the ring radius using different assumptions for the central gravitational field. These results are in a good agreement with the
observed values reported by  \citet{vanderLaan:2013qy} and \citet{Beifiori:2012}.

\begin{figure}[t!]
  \centering
  \includegraphics[scale=0.48]{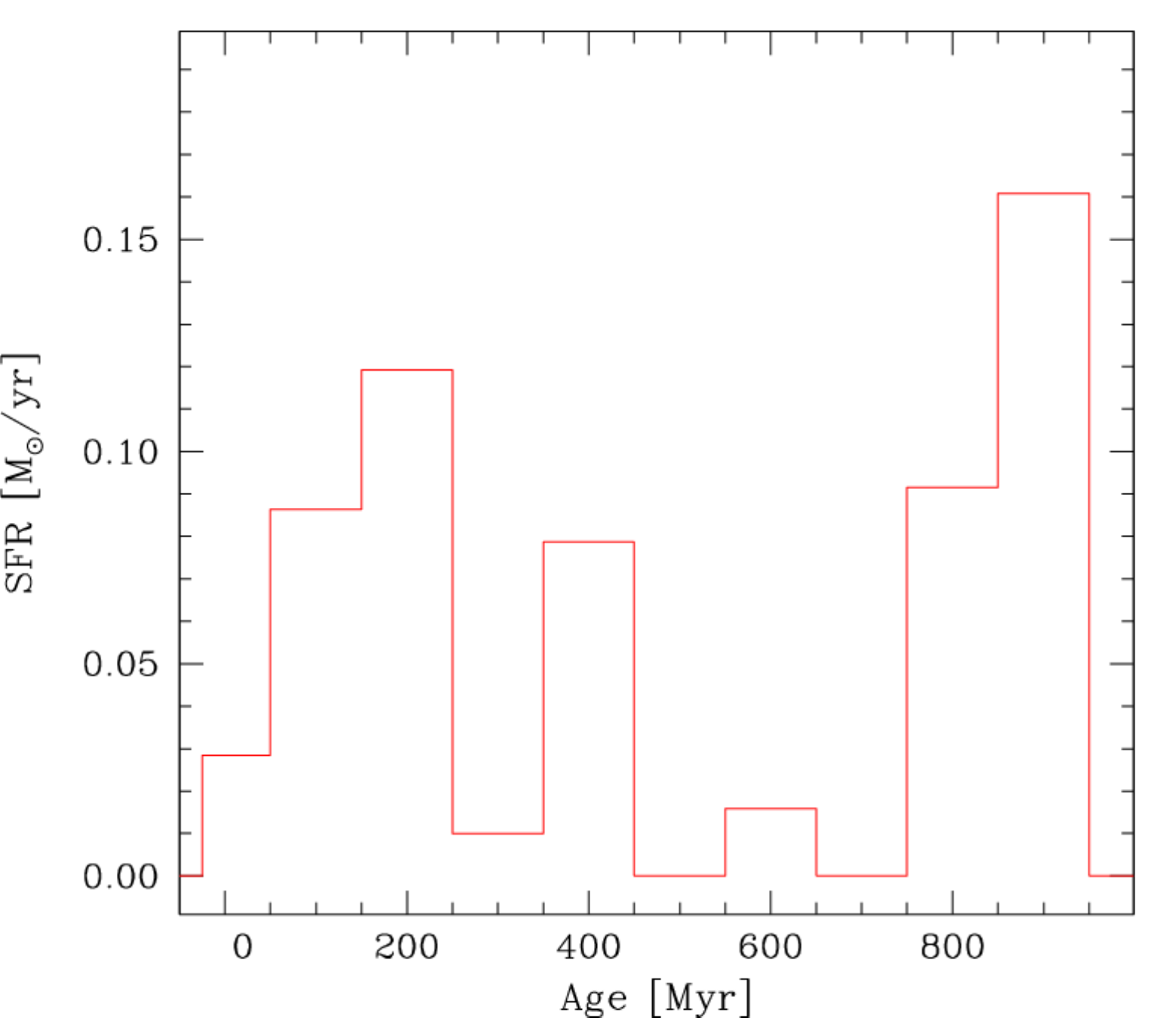}
   \caption{Star formation history in the circumnuclear ring of NGC 6951 \citep{vanderLaan:2013qy}.}
  \label{fig:16}
\end{figure}

\begin{figure}[t!]
  \centering
  \includegraphics[width=\columnwidth]{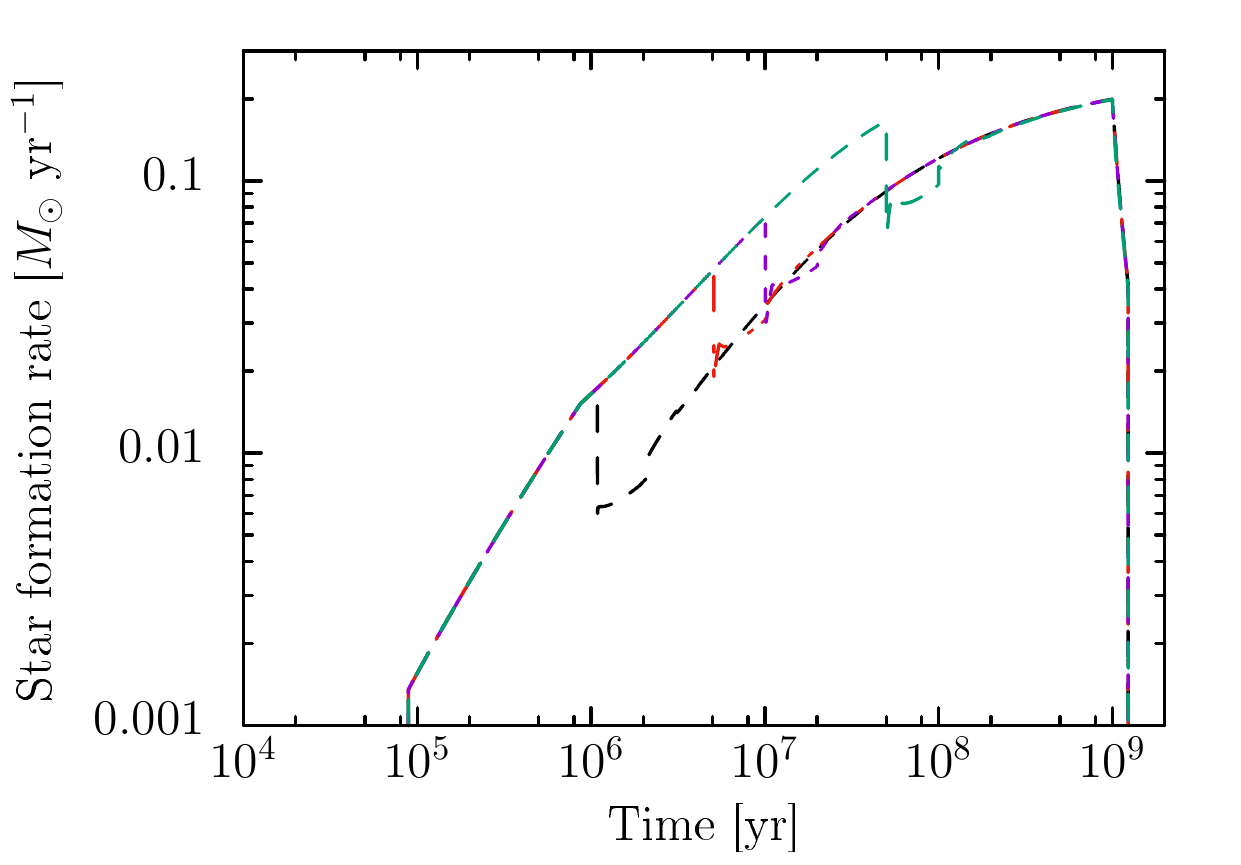}
%\resizebox{\columnwidth}{!}{\input{NGC6951-SN.tex}}
   \caption{Simulation of the  time evolution of the star formation rate (in the black hole field domination) of the starburst ring of  NGC 6951.
   Different coloured dashed-lines indicate the feedback from supernovae
   explosions assuming different lifetimes for the responsible massive stars: $10^{6}$ years (black), $5\times 10^{6}$ years (red), $10^{7}$ years (violet) and~
   $5\times~10^{7}$~years (green).}
  \label{fig:17}
\end{figure}

Figure \ref{fig:16} displays the  star formation rate history in the ring, where the stellar ages range from million to billion years. The simulated star formation rate versus 
time is shown in figure \ref{fig:17}. For producing all different coloured curves illustrated in figure \ref{fig:17},  we have employed the input parameters as mentioned above. 
The coloured curves correspond to the different lifetimes of massive stars exploding into supernovae.  All curves have the same final results for the masses and for the ring radius when 
employing the same assumptions on gravity as displayed in table \ref{tab:3}. Due to the delayed supernova feedback for larger lifetimes, in the figure we can see that 
the SFR initially remains higher~(>~0.02~$M_{\odot}$~$\rm{yr}^{-1}$)  until the feedback sets in. In general, such fluctuations in the SFR induced by supernova feedback are in good 
agreement with the observed star formation rate variation (figure \ref{fig:16}) by  \citet{vanderLaan:2013qy}, implying that our model can reproduce  the  observed star 
formation rate oscillating between 0 and~0.2~$M_{\odot}$~$\rm{yr}^{-1}$.

  \begin{table}[t!]
  \centering
  \caption{Results from simulations: final values of the physical properties of NGC 6951.}
  \label{tab:par}
 \vspace{0.3cm}
%  \begin{tabular}{ lp{1cm}lp{2.5cm}lp{1cm}lp{2cm}l }
   \begin{tabular}  {  m{1.9cm}  m{1.8cm} m{1.8cm}  m{1.7cm}   } 
%  \begin{tabular}{ lccccl }
    \hline
  	Source of the gravitational field& Final ring stellar mass $[10^8$ $M_{\odot}]$  & Ring radius [pc] & Final black hole mass $[10^7$ $ M_{\odot}]$ \\ 
	\hline
	%\noalign{\smallskip}
	%\multirow{2}{*}{Defenders} 
	%Observation &	$1.89\times10^{8}$ & 576 & $4.15\times10^{7}$ \\ 
	Central Black hole &	1.89 & 576 & 4.15 \\ 
	CND &	 1.79& 573 &4.38 \\
        Host galaxy& 1.93 & 611	&1.56 \\
       	\hline
  \end{tabular}
  \label{tab:3}
\end{table}

\section{Discussion and conclusions}

We have extended the co-evolution model of \citet{Wutschik:2013caa} by implementing the turbulence-regulated star 
formation model of \citet{Krumholz:2005mt}(model KM05). We have additionally studied the time evolution of the Mach numbers as well as the AGN 
luminosity. The main results of this work are the following:

\begin{itemize}
 \item [$\bullet$] The implementation of  the KM05 model produces moderate fluctuations in the black hole mass accretion rates and star formation rates 
 during the accretion phase rather than the strong fluctuations produced by the model EB10. The latter predicts over-accretion, i.e. the sonic speed becomes extremely 
 large compared to the local sound speed which produces, in consequence, extreme Mach numbers  up to $10^{4}$.
\end{itemize}

\begin{itemize}
 \item [$\bullet$] We have studied the limiting cases of the total star formation rate under different assumptions for the gravitational field: dominated by the central
 black hole, the CND or the host galaxy.  For each of these cases, our model is able to predict the growth of a SMBH from 10 million solar masses 
up to a few billions solar masses, with low (1 $M_{\odot}$ $\rm{yr}^{-1}$) and very large (1000 $M_{\odot}$ $\rm{yr}^{-1}$) mass supply rates 
respectively,  in agreement with earlier work by  \citet{Wutschik:2013caa}, \citet{Kawakatu:ka2009} and \citet{Volonteri2015ApJ}.
On the other hand, though that model EB10 produces an over-accretion, the final black hole masses are similar to the final values provided by model KM05  for gravity 
dominated by the CND. The final stellar masses generated by both models are also identical when compared at all field dominations.
\end{itemize}

\begin{itemize}
 \item [$\bullet$] An increased mass supply rate enhances black hole accretion rates, star formation rate, disk radius
final, black hole masses, final stellar masses as well as luminosities. However, we note that  large mass supply 
rates dampen the  level of turbulence and the competition of gas accretion during high accretion phases. This effect becomes 
more evident on the Mach number evolution when gravity is dominated by the central black hole. Large mass supply rates also significantly affect the disk radius; they lead  
to the growth of  the inner and outer radii, from parsec scale to a few kilo-parsec. For instance, to produce an SMBH of a 
billion solar masses, our model predicts a final inner radius of 10 pc and outer radius of 500 pc. Our model shows that large mass supply rates have not only a significant impact 
on the final black hole mass or black hole accretion rates but also on the order of magnitude of AGN luminosities. 
\end{itemize}
\begin{itemize}
 \item [$\bullet$] Since our model considers high and low accretion phases during the evolution of the system, we have also assumed 
that the time evolution of the AGN luminosities follows similar phases accordingly. Adopting the slim disk approximation during the high accretion phase our model predicts 
super-Eddington luminosities, whereas sub-Eddington luminosities are found during the low accretion 
phase when the standard thin disk approximation is applied. The latter  is reflected in the decrease of the luminosity  after the end of the mass supply. 
However, afterwards it seems that the central black hole keeps accreting at very low rates of
the order of $10^{-3}$ $M_{\odot}$ $\rm{yr}^{-1}$.
\end{itemize}

\begin{itemize}
 \item [$\bullet$] As a result of the  extended  supply time and increased mass supply rate, our model is able to predict an upper  limit on the AGN  luminosity  
 which is close to $10^{47}$ erg/s. AGN luminosities lower than $10^{47}$ erg/s are in agreement with \citet{Kawakatu:2008rg}  and \citet{Kawakatu:ka2009}. However, 
 the upper limit is reached at a different time than the cosmological age of the observed AGN by \citet{Mortlock:2011lo} and \citet{Pezzulli2016MNRAS} (with references therein). However, we do not know when the mass supply started in the observed systems, and of course the real mass supply evolution may be more complex, i.e. time dependent.  Despite this possible discrepancy with respect to time scales, our model succeeds in predicting the observed final black hole masses.
 \end{itemize}

\begin{itemize}
 \item [$\bullet$] We find that turbulence is a main driver of  angular momentum redistribution \citep{Balbus:Bal1998};\citep{Klessen:kle2010}, and particularly high values 
 of turbulence are found in the phase of high star formation rates and for the gravitational field dominated by the central black hole. We note that the turbulent velocity 
 of gas becomes larger not only during the high accretion phase but also when the gas supply is depleted. 
 While energy is injected by supernova explosions already at earlier stages of the accretion, the explosions occuring at the later stages are associated with the rise of the 
 turbulent velocity due to a decreasing density, which is reflected in the peaks of the Mach number in the late evolutionary stages.
  We emphasize here that the timescale of supernova feedback sets the timescale for fluctuations in both 
 the black hole accretion rate and star formation rate.
\end{itemize}

\begin{itemize}
 \item [$\bullet$] Our model does not take into account  any feedback from the AGN but if considered, it may indeed decrease the accretion
rates and  the  jet emission may trigger  star formation as investigated in MHD simulations by  \cite{MNR:MNR21479}.
\end{itemize}

\begin{itemize}
 \item [$\bullet$] While our model does neither assume a specific structure of the galaxy nor  take into account the interaction with the bar in NGC 6951,
 some quantities like the black hole mass  and the stellar mass can  still be reproduced. This suggests that these quantities may not strongly depend on 
 the choice of the gravitational potential, though further investigations may be necessary in the future.
\end{itemize}

\begin{acknowledgements}
We thank the anonymous referee for insightful comments regarding our manuscript. We are grateful to Tessel van der Laan for permission to include figure 
\ref{fig:16} in our manuscript. DRGS thanks for funding through Fondecyt regular (project code 1161247), through the ''Concurso Proyectos Internacionales de 
Investigaci\'on, Convocatoria 2015'' (project code PII20150171), trough ALMA-Conicyt (project code 31160001) and from the Chilean  BASAL  Centro  de  Excelencia 
en Astrof\'isica y Tecnolog\'ias Afines (CATA) grant FB-06/2007. WC  wishes to thank to M. Breuhaus for his encouragements and helpful discussions.
\end{acknowledgements}

%\bibliography{sample}

\begin{thebibliography}{62}
\expandafter\ifx\csname natexlab\endcsname\relax\def\natexlab#1{#1}\fi

\bibitem[{Abramowicz(2004)}]{Abramowicz:2004vi}
Abramowicz, M.~A. 2004, in {Conference on Growing Black Holes: Accretion in a
  Cosmological Context Garching, Germany, June 21-25, 2004}

\bibitem[{Abramowicz {et~al.}(1988)Abramowicz, Czerny, Lasota, \&
  Szuszkiewicz}]{Abramowicz:1988sp}
Abramowicz, M.~A., Czerny, B., Lasota, J.~P., \& Szuszkiewicz, E. 1988,
  Astrophys. J., 332, 646

\bibitem[{Armijo \& Pacheco(2011)}]{Armijo:2010md}
Armijo, M.~M. \& Pacheco, J.~F. 2011, Astron.Astrophys., 526, A146

\bibitem[{{Athanassoula}(1992)}]{Athanassoula1992MNRAS328A}
{Athanassoula}, E. 1992, \mnras, 259, 328

\bibitem[{{Balbus} \& {Hawley}(1998)}]{Balbus:Bal1998}
{Balbus}, S.~A. \& {Hawley}, J.~F. 1998, Reviews of Modern Physics, 70, 1

\bibitem[{{Begelman} \& {Shlosman}(2009)}]{Begelman09}
{Begelman}, M.~C. \& {Shlosman}, I. 2009, \apjl, 702, L5

\bibitem[{Beifiori {et~al.}(2012)Beifiori, Courteau, Corsini, \&
  Zhu}]{Beifiori:2012}
Beifiori, A., Courteau, S., Corsini, E., \& Zhu, Y. 2012, MNRAS., 419, 2497

\bibitem[{{Devecchi} {et~al.}(2010){Devecchi}, {Volonteri}, {Colpi}, \&
  {Haardt}}]{Devecchi10}
{Devecchi}, B., {Volonteri}, M., {Colpi}, M., \& {Haardt}, F. 2010, \mnras,
  409, 1057

\bibitem[{{Devecchi} {et~al.}(2012){Devecchi}, {Volonteri}, {Rossi}, {Colpi},
  \& {Portegies Zwart}}]{Devecchi12}
{Devecchi}, B., {Volonteri}, M., {Rossi}, E.~M., {Colpi}, M., \& {Portegies
  Zwart}, S. 2012, \mnras, 421, 1465

\bibitem[{Diamond-Stanic \& Rieke(2012)}]{DiamondStanic:2011md}
Diamond-Stanic, A.~M. \& Rieke, G.~H. 2012, Astrophys.J., 746, 168

\bibitem[{Dickel(1978)}]{Dickel:1978di}
Dickel, J. R.and~Rood, H.~J. 1978, Astrophys. J., 223, 391

\bibitem[{Elmegreen \& Burkert(2010)}]{Elmegreen:2009mv}
Elmegreen, B.~G. \& Burkert, A. 2010, Astrophys.J., 712, 294

\bibitem[{Faber \& Gallagher(1979)}]{Faber:1979fa}
Faber, S.~M. \& Gallagher, J.~S. 1979, Annual Review of Astronomy and
  Astrophysics, 17, 135

\bibitem[{Falcon-Barroso {et~al.}(2013)Falcon-Barroso, Ramos~Almeida, Boker,
  Schinnerer, Knapen, Lancon, Ryder, {et~al.}}]{falcon:2013}
Falcon-Barroso, J., Ramos~Almeida, C., Boker, T., {et~al.} 2013
  [\eprint[arXiv]{1311.2041}]

\bibitem[{Fan {et~al.}(2004)Fan, Hennawi, Richards, Strauss, Schneider, Donley,
  Young, Annis, Lin, Lampeitl, Lupton, Gunn, Knapp, Brandt, Anderson, Bahcall,
  Brinkmann, Brunner, Fukugita, Szalay, Szokoly, \& York}]{Fan2004}
Fan, X., Hennawi, J.~F., Richards, G.~T., {et~al.} 2004, The Astronomical
  Journal, 128, 515

\bibitem[{Fan {et~al.}(2006)Fan, Strauss, Richards, Hennawi, Becker, White,
  Diamond-Stanic, Donley, Jiang, Kim, Vestergaard, Young, Gunn, Lupton, Knapp,
  Schneider, Brandt, Bahcall, Barentine, Brinkmann, Brewington, Fukugita,
  Harvanek, Kleinman, Krzesinski, Long, Eric H.~Neilsen, Nitta, Snedden, \&
  Voges}]{Fan2006}
Fan, X., Strauss, M.~A., Richards, G.~T., {et~al.} 2006, The Astronomical
  Journal, 131, 1203

\bibitem[{{Ferrara} {et~al.}(2014){Ferrara}, {Salvadori}, {Yue}, \&
  {Schleicher}}]{Ferrara14}
{Ferrara}, A., {Salvadori}, S., {Yue}, B., \& {Schleicher}, D. 2014, \mnras,
  443, 2410

\bibitem[{Ferrarese \& Merritt(2000)}]{Ferrarese:2000se}
Ferrarese, L. \& Merritt, D. 2000, Astrophys.J., 539, L9

\bibitem[{Gaibler {et~al.}(2012)Gaibler, Khochfar, Krause, \&
  Silk}]{MNR:MNR21479}
Gaibler, V., Khochfar, S., Krause, M., \& Silk, J. 2012, Monthly Notices of the
  Royal Astronomical Society, 425, 438

\bibitem[{Gebhardt {et~al.}(2000)Gebhardt, Bender, Bower, Dressler, Faber,
  {et~al.}}]{Gebhardt:2000fk}
Gebhardt, K., Bender, R., Bower, G., {et~al.} 2000, Astrophys.J., 539, L13

\bibitem[{Haring \& Rix(2004)}]{Haring:2004hr}
Haring, N. \& Rix, H.-W. 2004, Astrophys.J., 604, L89

\bibitem[{Hsieh {et~al.}(2011)Hsieh, Matsushita, Liu, Ho, Oi,
  {et~al.}}]{Hsieh:2011ki}
Hsieh, P.-Y., Matsushita, S., Liu, G., {et~al.} 2011, Astrophys.J., 736, 129

\bibitem[{Kawakatu \& Wada(2008)}]{Kawakatu:2008rg}
Kawakatu, N. \& Wada, K. 2008, Astrophys.J., 681, 73

\bibitem[{Kawakatu \& Wada(2009)}]{Kawakatu:ka2009}
Kawakatu, N. \& Wada, K. 2009, The Astrophysical Journal, 706, 676

\bibitem[{Klessen \& Hennebelle(2010)}]{Klessen:kle2010}
Klessen, R.~S. \& Hennebelle, P. 2010, Astron.Astrophys., 520, A17

\bibitem[{Krumholz \& McKee(2005)}]{Krumholz:2005mt}
Krumholz, M.~R. \& McKee, C.~F. 2005, Astrophys.J., 630, 250

\bibitem[{Latif {et~al.}(2013{\natexlab{a}})Latif, Schleicher, Schmidt, \&
  Niemeyer}]{Latif:2013dua}
Latif, M., Schleicher, D., Schmidt, W., \& Niemeyer, J. 2013{\natexlab{a}}
  [\eprint[arXiv]{1304.0962}]

\bibitem[{Latif {et~al.}(2013{\natexlab{b}})Latif, Schleicher, Schmidt, \&
  Niemeyer}]{Latif:2013pyq}
Latif, M., Schleicher, D., Schmidt, W., \& Niemeyer, J. 2013{\natexlab{b}}
  [\eprint[arXiv]{1309.1097}]

\bibitem[{{Latif} \& {Ferrara}(2016)}]{Latif2016}
{Latif}, M.~A. \& {Ferrara}, A. 2016, ArXiv e-prints
  [\eprint[arXiv]{1605.07391}]

\bibitem[{{Lena} {et~al.}(2015){Lena}, {Robinson}, {Storchi-Bergman},
  {Schnorr-M{\"u}ller}, {Seelig}, {Riffel}, {Nagar}, {Couto}, \&
  {Shadler}}]{Lena2015ApJ}
{Lena}, D., {Robinson}, A., {Storchi-Bergman}, T., {et~al.} 2015, \apj, 806, 84

\bibitem[{Lenc \& Tingay(2009)}]{Lenc:2009le}
Lenc, E. \& Tingay, S.~J. 2009, The Astronomical Journal, 137, 537

\bibitem[{{Lodato} \& {Natarajan}(2007)}]{Lodato07}
{Lodato}, G. \& {Natarajan}, P. 2007, \mnras, 377, L64

\bibitem[{Magorrian {et~al.}(1998)Magorrian, Tremaine, Richstone, Bender,
  Bower, {et~al.}}]{Magorrian:1997hw}
Magorrian, J., Tremaine, S., Richstone, D., {et~al.} 1998, Astron.J., 115, 2285

\bibitem[{Marconi \& Hunt(2003)}]{Marconi:2003hj}
Marconi, A. \& Hunt, L.~K. 2003, Astrophys.J., 589, L21

\bibitem[{Merritt(1999)}]{Merritt:1999ry}
Merritt, D. 1999 [\eprint[arXiv]{astro-ph/9910546}]

\bibitem[{Mortlock {et~al.}(2011)Mortlock, Warren, Venemans, Patel, Hewett,
  McMahon, Simpson, Theuns, Gonz\'ales-Solares, Adamson, Dye, Hambly, Hirst,
  Irwin, Kuiper, Lawrence, \& Roettgering}]{Mortlock:2011lo}
Mortlock, D.~J., Warren, S.~J., Venemans, B.~P., {et~al.} 2011, {Nature}

\bibitem[{Perez {et~al.}(1999)Perez, Marquez, Marrero, Durret,
  Gonzalez~Delgado, {et~al.}}]{Perez:1999je}
Perez, E., Marquez, I., Marrero, I., {et~al.} 1999
  [\eprint[arXiv]{astro-ph/9909495}]

\bibitem[{{Pezzulli} {et~al.}(2016){Pezzulli}, {Valiante}, \&
  {Schneider}}]{Pezzulli2016MNRAS}
{Pezzulli}, E., {Valiante}, R., \& {Schneider}, R. 2016, \mnras, 458, 3047

\bibitem[{Pringle(1981)}]{Pringle:1981ds}
Pringle, J. 1981, Ann.Rev.Astron.Astrophys., 19, 137

\bibitem[{Rees(1984)}]{Rees:1984tin}
Rees, M.~J. 1984, Annual review of astronomy and astrophysics, 471

\bibitem[{Regan \& Teuben(2003)}]{Regan2003}
Regan, M.~W. \& Teuben, P.~J. 2003, Astrophys. J., 582, 723

\bibitem[{{Rice}(2016)}]{Rice:2016ri}
{Rice}, K. 2016, \pasa, 33, e012

\bibitem[{{Sani} {et~al.}(2012){Sani}, {Davies}, {Sternberg},
  {Graci{\'a}-Carpio}, {Hicks}, {Krips}, {Tacconi}, {Genzel}, {Vollmer},
  {Schinnerer}, {Garc{\'{\i}}a-Burillo}, {Usero}, \& {Orban de
  Xivry}}]{Sani:2012sani}
{Sani}, E., {Davies}, R.~I., {Sternberg}, A., {et~al.} 2012, MNRAS, 424, 1963

\bibitem[{{Schleicher} {et~al.}(2013){Schleicher}, {Palla}, {Ferrara}, {Galli},
  \& {Latif}}]{Schleicher13}
{Schleicher}, D.~R.~G., {Palla}, F., {Ferrara}, A., {Galli}, D., \& {Latif}, M.
  2013, \aap, 558, A59

\bibitem[{{Schleicher} {et~al.}(2010{\natexlab{a}}){Schleicher}, {Spaans}, \&
  {Glover}}]{Schleicher:2010apj}
{Schleicher}, D.~R.~G., {Spaans}, M., \& {Glover}, S.~C.~O. 2010{\natexlab{a}},
  \apjl, 712, L69

\bibitem[{{Schleicher} {et~al.}(2010{\natexlab{b}}){Schleicher}, {Spaans}, \&
  {Klessen}}]{Schleicher:2010AA}
{Schleicher}, D.~R.~G., {Spaans}, M., \& {Klessen}, R.~S. 2010{\natexlab{b}},
  \aap, 513, A7

\bibitem[{{Schnorr-M{\"u}ller}
  {et~al.}(2014{\natexlab{a}}){Schnorr-M{\"u}ller}, {Storchi-Bergmann},
  {Nagar}, \& {Ferrari}}]{Schnorr2014MNRASa}
{Schnorr-M{\"u}ller}, A., {Storchi-Bergmann}, T., {Nagar}, N.~M., \& {Ferrari},
  F. 2014{\natexlab{a}}, \mnras, 438, 3322

\bibitem[{{Schnorr-M{\"u}ller}
  {et~al.}(2014{\natexlab{b}}){Schnorr-M{\"u}ller}, {Storchi-Bergmann},
  {Nagar}, {Robinson}, {Lena}, {Riffel}, \& {Couto}}]{Schnorr2014MNRASb}
{Schnorr-M{\"u}ller}, A., {Storchi-Bergmann}, T., {Nagar}, N.~M., {et~al.}
  2014{\natexlab{b}}, \mnras, 437, 1708

\bibitem[{{Schnorr-M{\"u}ller} {et~al.}(2016){Schnorr-M{\"u}ller},
  {Storchi-Bergmann}, {Robinson}, {Lena}, \& {Nagar}}]{Schnorr2016MNRAS}
{Schnorr-M{\"u}ller}, A., {Storchi-Bergmann}, T., {Robinson}, A., {Lena}, D.,
  \& {Nagar}, N.~M. 2016, \mnras, 457, 972

\bibitem[{Shakura \& Sunyaev(1973)}]{Shakura:1972te}
Shakura, N.~I. \& Sunyaev, R.~A. 1973, Astron. Astrophys., 24, 337

\bibitem[{Spaans \& Meijerink(2008)}]{Spaans2008}
Spaans, M. \& Meijerink, R. 2008, The Astrophysical Journal Letters, 678, L5

\bibitem[{{Storchi-Bergmann} {et~al.}(2010){Storchi-Bergmann}, {Lopes},
  {McGregor}, {Riffel}, {Beck}, \& {Martini}}]{Storchi2010MNRAS}
{Storchi-Bergmann}, T., {Lopes}, R.~D.~S., {McGregor}, P.~J., {et~al.} 2010,
  \mnras, 402, 819

\bibitem[{{Storchi-Bergmann} {et~al.}(2009){Storchi-Bergmann}, {McGregor},
  {Riffel}, {Sim{\~o}es Lopes}, {Beck}, \& {Dopita}}]{Storchi2009MNRAS}
{Storchi-Bergmann}, T., {McGregor}, P.~J., {Riffel}, R.~A., {et~al.} 2009,
  \mnras, 394, 1148

\bibitem[{Toomre(1964)}]{Toomre:1964zx}
Toomre, A. 1964, Astrophys. J., 139, 1217

\bibitem[{{Trenti} \& {Stiavelli}(2007)}]{Trenti07}
{Trenti}, M. \& {Stiavelli}, M. 2007, \apj, 667, 38

\bibitem[{van~der Laan {et~al.}(2013)van~der Laan, Schinnerer, Emsellem, Hunt,
  McDermid, {et~al.}}]{vanderLaan:2013qy}
van~der Laan, T., Schinnerer, E., Emsellem, E., {et~al.} 2013
  [\eprint[arXiv]{1301.2621}]

\bibitem[{{Venemans} {et~al.}(2012){Venemans}, {McMahon}, {Walter}, {Decarli},
  {Cox}, {Neri}, {Hewett}, {Mortlock}, {Simpson}, \&
  {Warren}}]{Venemans2012ApJ}
{Venemans}, B.~P., {McMahon}, R.~G., {Walter}, F., {et~al.} 2012, \apjl, 751,
  L25

\bibitem[{Volonteri {et~al.}(2015)Volonteri, Silk, \& Dubus}]{Volonteri2015ApJ}
Volonteri, M., Silk, J., \& Dubus, G. 2015, Astrophys. J., 804, 148

\bibitem[{{Watarai} {et~al.}(2000){Watarai}, {Fukue}, {Takeuchi}, \&
  {Mineshige}}]{Watarai2000}
{Watarai}, K.-y., {Fukue}, J., {Takeuchi}, M., \& {Mineshige}, S. 2000, \pasj,
  52, 133

\bibitem[{{Willott} {et~al.}(2003){Willott}, {McLure}, \&
  {Jarvis}}]{Willot2003ApJ}
{Willott}, C.~J., {McLure}, R.~J., \& {Jarvis}, M.~J. 2003, \apjl, 587, L15

\bibitem[{{Wise} {et~al.}(2008){Wise}, {Turk}, \& {Abel}}]{Wise08}
{Wise}, J.~H., {Turk}, M.~J., \& {Abel}, T. 2008, \apj, 682, 745

\bibitem[{Wutschik {et~al.}(2013)Wutschik, Schleicher, \&
  Palmer}]{Wutschik:2013caa}
Wutschik, S., Schleicher, D. R.~G., \& Palmer, T.~S. 2013, Astron.Astrophys.,
  560, A34

\end{thebibliography}
%\bibliographystyle{aa}

%\appendix

%\section{Additional info}

\end{document}